\documentclass[12pt]{article}%
\usepackage{amsmath}
\usepackage{amsfonts}
\usepackage{amssymb}
\usepackage{graphicx}%
\providecommand{\U}[1]{\protect\rule{.1in}{.1in}}
\begin{document}
\begin{titlepage}
%\begin{flushright}
%Preliminary
%\end{flushright}
\begin{center}
\LARGE
{\bf
Quantum Energy Teleportation:\\
An Introductory Review
}
\end{center}
%\begin{center}
%\large{
%For Kusatsu Winter Workshop on\\
%Non-Equilibrium Fluctuation in Physics and Informatics,\\
%7-10 March, 2011, Gunma, Japan.
%}
%\end{center}
\ \\
\ \\
\begin{center}
\Large{
Masahiro Hotta
}\\
\ \\
{\it
Graduate School of Science, Tohoku University,\\
Sendai 980-8578, Japan\\
hotta@tuhep.phys.tohoku.ac.jp
}
\end{center}
\begin{abstract}
The development of techniques for manipulation of quantum information has opened the door to a variety of protocols for accomplishing unprecedented tasks.
In particular, a new protocol of quantum teleportation was proposed in 2008 to achieve effective energy transportation simply via local operations and classical communication
---without breaking any known physical laws. This is achieved
by extracting zero-point energy of entangled many-body systems by local operations dependent on information obtained by a distant measurement.
The concept is reviewed from an introductory viewpoint herein.
\end{abstract}
\end{titlepage}

\bigskip

\section{Introduction}

\ \newline

Together with spacetime, matter and information constitute the great building
blocks of the Universe. Matter is generally known to comprise the elementary
particles such as electrons and photons. But more precisely speaking, the
concept encompasses all conserved physical quantities such as energy and
electric charge carried by local excitations of elementary-particle fields.
Information, however, represents the quantum information carried by matter.
All quantum information is consolidated by the quantum state of the fields. A
significant essential property of matter is its sameness at a very deep
bottom. For example, the properties of an electron at one position are
indistinguishable from those of another electron at a different position: they
have the same mass, the same spin, and the same electric charge. Quantum
mechanics does not distinguish between two particles corresponding to the same
field. Then from whence cometh the distinguishing characteristics of
individuals and the objects surrounding us? They stem from the full
information imprinted on common matter, called the quantum field, that is, the
quantum state. In 1982, Wootters and Zurek discovered a remarkable theorem
about quantum states \cite{wz}. Their theorem shows that general quantum
states prohibit their cloning. In contrast to classical information, we cannot
make indistinguishable copies of quantum information. In this sense, quantum
information is one of the most profound concepts about identity.

Taking into account the above consideration, it is possible to argue that
transportation of a quantum state is equivalent to transportation of an object
itself with individual characteristics. In quantum mechanics, the
transportation of quantum states can be achieved simply by local operations
and classical communication (LOCC for short). This protocol was proposed in
1993 and named quantum teleportation \cite{QT}. Why is the protocol called
'teleportation'? To illustrate this reason concretely, let us consider the
protocol with qubits. Alice and Bob stay at different positions and share a
Bell pair of two qubits $A$ and $B$ in a state%

\[
|E_{0}\rangle_{AB}=\frac{1}{\sqrt{2}}\left(  |+\rangle_{A}|+\rangle
_{B}+|-\rangle_{A}|-\rangle_{B}\right)  ,
\]
where $|+\rangle$ $(|-\rangle)$ is the up (down) state of the third Pauli
operator $\sigma_{3}(=|+\rangle\langle+|-|-\rangle\langle-|)$. Alice also has
another qubit $A^{\prime}$ in a unknown state $|\psi\rangle$. The state of
these three qubits can be calculated as%

\begin{equation}
|\psi\rangle_{A^{\prime}}\otimes|E_{0}\rangle_{AB}=\frac{1}{2}\sum_{\alpha
=0}^{3}|E_{\alpha}\rangle_{A^{\prime}A}\otimes\sigma_{\alpha B}|\psi
\rangle_{B}, \label{e1}%
\end{equation}
where the orthogonal Bell states $|E_{\alpha}\rangle_{A^{\prime}A}$ are given
by $\sigma_{\alpha A^{\prime}}|E_{0}\rangle_{A^{\prime}A}$, and $\sigma
_{\alpha A^{\prime}}~(\sigma_{\alpha B})$ is the Pauli operator $\sigma
_{\alpha}$ for $A^{\prime}$ $(B)$ with its 0-th component $\sigma_{0}=I$. At
time $t=t_{m}$, Alice performs a Bell measurement to identify which
$|E_{\alpha}\rangle_{A^{\prime}A}$ is realized for the composite system of
$A^{\prime}\,$and $A$. The output is two-bit information of $\alpha
(=0,1,2,3)$. Each emergence probability of $\alpha$ is the same and equal to
$1/4$. Because of so-called wavefunction collapse in quantum measurement, the
system in the state $|\psi\rangle_{A^{\prime}}\otimes|E_{0}\rangle_{AB}$ jumps
instantaneously into a different state $|E_{\alpha}\rangle_{A^{\prime}%
A}\otimes\sigma_{\alpha B}|\psi\rangle_{B}$ corresponding to the measurement
result $\alpha$. Very surprisingly, the state of $B$ becomes a pure state
$\sigma_{\alpha}|\psi\rangle$ and acquires nontrivial dependence on the input
state $|\psi\rangle$. This means that $B$ suddenly gets information about
$|\psi\rangle$ at the moment of a distant measurement by Alice. In this sense,
the quantum information is 'teleported' from Alice to Bob at $t=t_{m}$. The
instantaneous state change at $t=t_{m}$ is depicted in figure 1.%
%TCIMACRO{\FRAME{dtbpFU}{3.0649in}{2.3073in}{0pt}{\Qcb{\QTR{tiny}{Figure 1:
%Instantaneous state change after local measurement by Alice in the
%conventional quantum teleportation protocol with qubits.}}}{}{fig1.eps}%
%width 3.0649in;  height 2.3073in;  depth 0pt;  original-width 7.2627in;
%original-height 5.4518in;  cropleft "0";  croptop "1";  cropright "1";
%filename 'FIG-arXiv-QET-REVIEW/EPS/fig1.eps';file-properties "XNPEU";}}}%
%BeginExpansion
\begin{center}
\includegraphics[
height=2.3073in,
width=3.0649in
]%
{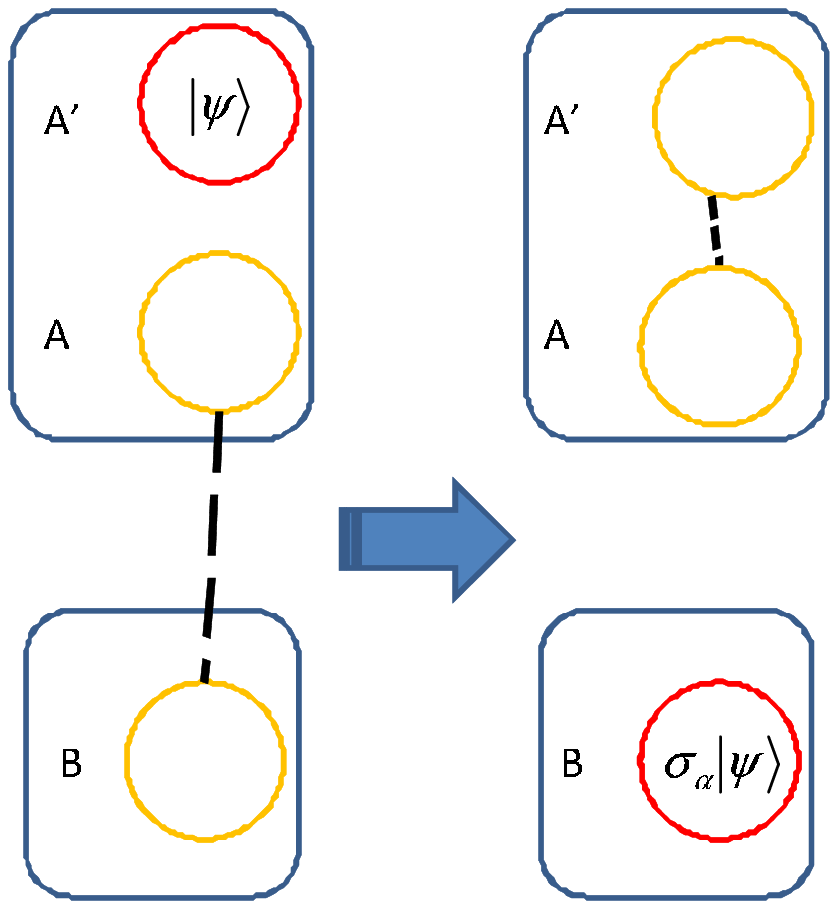}%
\\
{\protect\tiny Figure 1: Instantaneous state change after local measurement by
Alice in the conventional quantum teleportation protocol with qubits.}%
\end{center}
%EndExpansion
After the measurement, Alice announces the result $\alpha$ to Bob via a
classical channel such as a telephone. Though the post-measurement state
$\sigma_{\alpha}|\psi\rangle$ of $B$ is different from the original state
$|\psi\rangle$, Bob can transform it to the correct one by locally performing
a unitary operation $\sigma_{\alpha}^{-1}$ on $B$ at the arrival time
$t=t_{op}$ of the information about $\alpha$. In order to avoid possible
misunderstanding of the concept of teleportation in physics, a comment must be
added here: Bob only knows which state of $\sigma_{\alpha}|\psi\rangle$
$(\alpha=0\sim3)$ is realized for $B$ after receiving information about the
measurement from Alice. The speed of this classical communication between them
never exceeds the speed of light and, therefore, causality is strictly retained.

The protocol of quantum teleportation \cite{QT} is really interesting.
However, it is not sufficient to teleport energy by itself. Transfer of an
excited state to a distant point requires preparation in advance of the same
amount of energy of the state at the point. If we do not have enough energy
around the distant point, the protocol never works. For example, let us
imagine that Alice sends to Bob the spin-up state of $\sigma_{3}$ of a qubit
in an external uniform magnetic field parallel to the z axis. For the
teleportation, they must share two qubits in a Bell state. The Hamiltonian of
each qubit is given by $H_{b}=b\sigma_{3}$ with a positive constant $b$. Note
that, in the Bell state, Bob's qubit has zero energy on average. After the
state teleportation, the energy of Bob's qubit increases to $b$ on average
because the teleported state is the up state. Because Bob's operation in the
protocol is local, it is clear that $b$ of the averaged energy must be
provided by an external operation device of Bob with a battery, for instance,
to drive it. During one round of the protocol, the energy of the battery
decreases by $b$ on average. If Bob does not have energy source like this
battery, the up-state teleportation does not succeed. On the other hand, if
the down state is teleported to Bob, Bob's qubit loses $b$ of energy on
average during his operation. Then the operation device receives $b$ of the
averaged energy as a work done by his qubit. Thus the down-state teleportation
may be accomplished even if Bob does not have energy sources. However, it
should be noticed that the averaged energy gain $b$ was originally available
for Bob without using the teleportation. Before the operation, Bob's qubit was
\textit{already excited} in a Bell state storing$\ b$ of energy, on average,
larger than that of the spin-down ground state. Bob's qubit merely has
disgorged the surplus energy due to the transition into the ground state.
Therefore, in this protocol, available energy for Bob moves around the region
of Bob \textit{without any increase of its total amount}. No energy is
teleported in this case. Then do the known laws of physics truly allow energy
teleportation? Can we teleport an object with energy to a zero-energy
local-vacuum region? Amazingly, the answer is yes---in principle. Energy can
be effectively transported simply using local operations and classical
communication, just like in the usual quantum teleportation protocol. In
quantum mechanics, we can generate quantum states containing a spatial region
with negative energy density of quantum fields \cite{bd}. Thus, even if we
have zero energy in a region where an object is going to be teleported, its
energy can be extracted from the vacuum fluctuation of quantum fields,
generating negative energy density around there. This can be attained by using
a local squeezing operation dependent on the result of a measurement at the
starting point of the teleportation. Of course, local energy conservation and
all the other physical laws are not violated in the energy teleportation. The
protocols, called quantum energy teleportation (QET for short), were first
proposed by this author in 2008. QET can be implemented, at least
theoretically, to various physical systems, including spin chains
\cite{1}-\cite{3}, cold trapped ions \cite{4}, harmonic chains \cite{5}, and
quantum fields \cite{6}-\cite{8}. Besides, it has been recently presented that
QET would be experimentally verified by using quantum Hall edge currents
\cite{YIH}. Herein, we reviewed the QET protocols from an introductory viewpoint.

The QET mechanism has various links to other research fields in physics. First
of all, future QET technology is expected to achieve rapid energy distribution
without thermal decoherence inside quantum devices. Because it is not energy
but classical information that is sent to the distant point, no heat is
generated in the energy transport channel during the short time period of QET
protocols. This aspect will assist in the development of quantum computers.
QET also has a close relation to a local-cooling problem, as is explained in
section 4. A measurement on a subsystem of a ground-state many-body system
often breaks entanglement among the subsystems. This measurement process is
accompanied by energy infusion to the system because the post-measurement
state is not the ground state but instead an excited state. Here we are able
to pose an interesting question: Soon after the energy infusion, is it
possible to extract all the infused energy using only local operations on the
measured subsystem? The answer is no, because, from information theory,
residual energy is unavoidable for this local-cooling process \cite{1}. The
residual energy is lower bounded by the total amount of energy that can be
teleported to other subsystems by using the measurement information. The
quantum local cooling and QET expose a new aspect of quantum Maxwell's demon
\cite{demon}, who 'watches' quantum fluctuations in the ground state. The
amount of teleported energy depends nontrivially on entanglement in the ground
state of a many-body system. Hence, QET analyses are also expected to shed new
light on complicated entanglement in condensed matter physics and to deepen
our understanding of phase transition at zero temperature, which has been
recently discussed using the entanglement \cite{pt}. Moreover, QET provides a
new method extracting energy from black holes \cite{8}: Outside a black hole,
we perform a measurement of quantum fields and obtain information about the
quantum fluctuation. Then positive-energy wave packets of the fields are
generated during the measurement and fall into the black hole. Even after
absorption of the wave packets by the black hole, we can retrieve a part of
the absorbed energy outside the horizon by using QET. This energy extraction
yields a decrease in the horizon area, which is proportional to the entropy of
the black hole. However, if we accidentally lose the measurement information,
we cannot extract energy anymore. The black-hole entropy is unable to
decrease. Therefore, the obtained measurement information has a very close
connection with the black hole entropy. This line of argument is expected to
lead to further understanding of the origin of black hole entropy, which is
often discussed in string theory \cite{string}.

The present review is organized as follows: Section 2 presents an elementary
description of the QET mechanism to allow the reader to capture the essence of
the concept. Section 3 then introduces the most simple example of QET. In
section 4, the general theory of QET is constructed for one-dimensional
discrete chain models. In section 5, QET with a relativistic quantum field in
one dimension is analyzed. The summary and some comments are provided in the
last section.

\ 

\bigskip

\bigskip

\section{Capturing the Essence of QET Mechanism}

~

In this section, an elementary intuitive explanation of QET is presented to
allow the reader to capture the essence of the mechanism. More rigorous
analyses follow in the later sections. From an operational viewpoint, QET
appears to be a kind of scientific magic trick. Let us first imagine a magic
trick using two separate \textit{empty} boxes A and B performed by Alice and
Bob. Alice infuses some amount of energy to A. Then a secret trick begins to
work inside A (figure 2).%
%TCIMACRO{\FRAME{dtbpFU}{3.0649in}{2.3073in}{0pt}{\Qcb{\QTR{tiny}{Figure 2:
%First step of the QET magic trick. Some energy is inputted into the empty box
%A.}}}{}{fig2.eps}{\special{ language "Scientific Word";  type "GRAPHIC";
%width 3.0649in;  height 2.3073in;  depth 0pt;  original-width 7.2627in;
%original-height 5.4518in;  cropleft "0";  croptop "1";  cropright "1";
%filename 'FIG-arXiv-QET-REVIEW/EPS/fig2.eps';file-properties "XNPEU";}}}%
%BeginExpansion
\begin{center}
\includegraphics[
height=2.3073in,
width=3.0649in
]%
{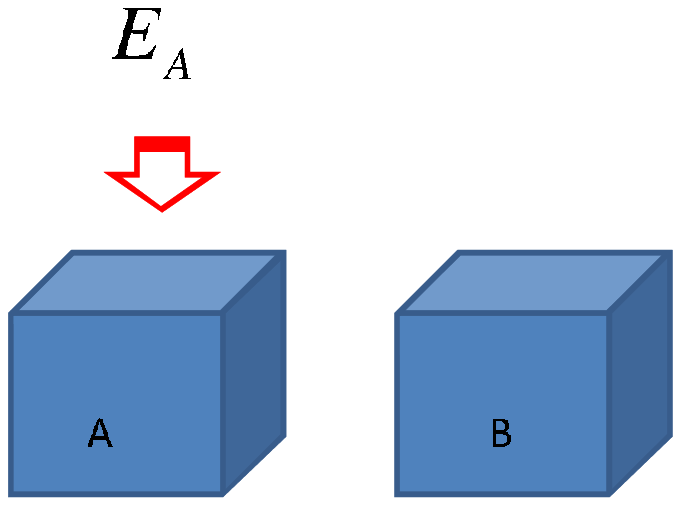}%
\\
{\protect\tiny Figure 2: First step of the QET magic trick. Some energy is
inputted into the empty box A.}%
\end{center}
%EndExpansion%
%TCIMACRO{\FRAME{dtbpFU}{3.0649in}{2.3073in}{0pt}{\Qcb{\QTR{tiny}{Figure 3:
%Second step of the QET magic trick, the abracadabra spell, which is a sequence
%of 0 and 1, is outputted from A. Then Alice announces it to Bob who is in
%front of B.}}}{}{fig3.eps}{\special{ language "Scientific Word";
%type "GRAPHIC";  maintain-aspect-ratio TRUE;  display "USEDEF";
%valid_file "F";  width 3.0649in;  height 2.3073in;  depth 0pt;
%original-width 7.2627in;  original-height 5.4518in;  cropleft "0";
%filename 'FIG-arXiv-QET-REVIEW/EPS/fig3.eps';file-properties "XNPEU";}}}%
%BeginExpansion
\begin{center}
\includegraphics[
height=2.3073in,
width=3.0649in
]%
{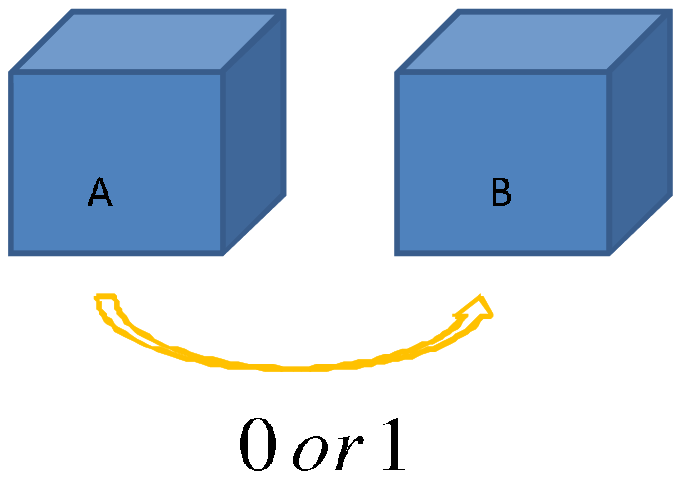}%
\\
{\protect\tiny Figure 3: Second step of the QET magic trick, the abracadabra
spell, which is a sequence of 0 and 1, is outputted from A. Then Alice
announces it to Bob who is in front of B.}%
\end{center}
%EndExpansion%
%TCIMACRO{\FRAME{dtbpFU}{3.0649in}{2.3073in}{0pt}{\Qcb{\QTR{tiny}{Figure 4:
%Third step of the QET magic trick. Bob inputs the abracadabra spell to B,
%which then begins to undergo an internal process that finally disgorges
%energy, even though B contained nothing at first.}}}{}{fig4.eps}%
%width 3.0649in;  height 2.3073in;  depth 0pt;  original-width 7.2627in;
%original-height 5.4518in;  cropleft "0";  croptop "1";  cropright "1";
%filename 'FIG-arXiv-QET-REVIEW/EPS/fig4.eps';file-properties "XNPEU";}}}%
%BeginExpansion
\begin{center}
\includegraphics[
height=2.3073in,
width=3.0649in
]%
{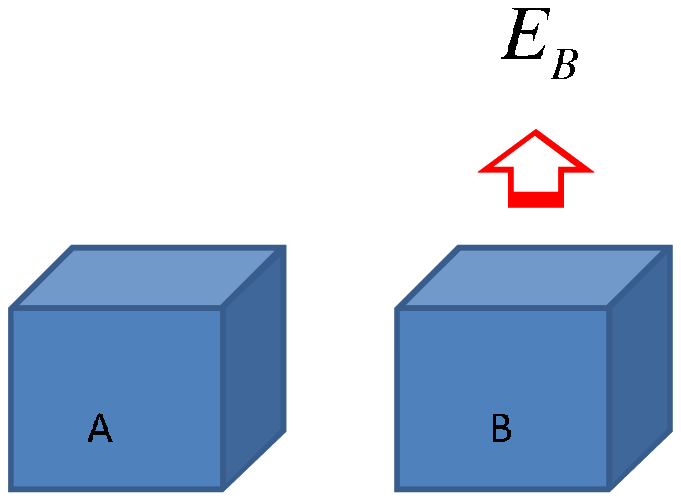}%
\\
{\protect\tiny Figure 4: Third step of the QET magic trick. Bob inputs the
abracadabra spell to B, which then begins to undergo an internal process that
finally disgorges energy, even though B contained nothing at first.}%
\end{center}
%EndExpansion
After a brief while, a magic spell, \textit{abracadabra}, which is a bit
number (0 or 1) in this case, is outputted from A. Then Alice announces this
information to Bob in front of B (figure 3) and Bob inputs the
\textit{abracadabra} to B. B begins some internal process and finally
disgorges energy, even though B contained \textit{nothing} at first (figure
4). Certainly this looks like energy teleportation. QET is able to achieve
this magic using quantum systems. Then what is the point of this QET magic?
The point is related to a question: What is \textit{nothing}? In quantum
theory, nothing means the ground state of the system, that is, the eigenstate
of total Hamiltonian corresponding to the minimum eigenvalue. For quantum
fields in particular, nothing means the vacuum state. It is a very surprising
fact of quantum mechanics that nonvanishing zero-point fluctuations exist even
in the vacuum state as \textit{nothing}.

What is zero-point fluctuation? In classical mechanics, a physical system is
completely frozen with no motion possible in its minimum energy state. For
example, let us consider a harmonic oscillator composed of a pendulous spring
attached to a ball of mass $m$. Assuming that the spring force at a position
$X$ is given by $F=-m\omega^{2}X$ with angular frequency $\omega$ of simple
harmonic oscillation, its Hamiltonian is given by%

\[
H=\frac{1}{2m}P^{2}+\frac{1}{2}m\omega^{2}X^{2}.
\]
In classical theory, both the position $X$ and momentum $P$ of the ball take
definite values. The ball is at rest at position $X=0$ and momentum $P=0$ in
the minimum energy state. However, in quantum theory, position and momentum
cannot simultaneously be fixed to arbitrary precision even in the ground state
according to Heisenberg's uncertainty principle: $\Delta X\Delta P\geq\hbar
/2$. Here, $\Delta X$ ($\Delta P$) is the quantum uncertainty in the position
(momentum) of the ball. Zero-point fluctuation is the random motion induced by
this quantum uncertainty. The minimum energy $E_{g}$ can be roughly estimated
by minimization of%

\[
E_{g}=\frac{1}{2m}\Delta P^{2}+\frac{1}{2}m\omega^{2}\Delta X^{2}%
\]
with $\Delta X\Delta P=\hbar/2$. This yields the following estimation:
\begin{equation}
E_{g}=\frac{1}{2}\hbar\omega. \label{e01}%
\end{equation}
It is well known that rigorous derivation of the ground-state energy also
gives the same result in Eq. (\ref{e01}). The ground-state energy is called
the zero-point energy. This simple example exposes that zero-point fluctuation
is capable of carrying nonzero energy. And it is not only the harmonic
oscillator but also other general interacting many-body systems that have
zero-point fluctuation with nonvanishing energy in the ground state. Each
subsystem is fluctuating with nonzero energy density in the ground state.
%TCIMACRO{\FRAME{dtbpFU}{3.0649in}{2.3073in}{0pt}{\Qcb{\QTR{tiny}{Figure 5:
%Zero-point fluctuation of a boson field in one dimension is schematically
%depicted. The vertical line implies amplitude of (coarse-grained) field
%fluctuation. The horizontal line describes spatial coordinate x. Mathematical
%description of the quantum fluctuation is a superposition of various
%configuration states. This situation is simplified in the figure and only two
%different configurations of the fluctuation are exhibited by red and blue
%broken lines. They fluctuate at a typical amplitude order fixed by the quantum
%uncertainty relation.}}}{}{fig5.eps}{\special{ language "Scientific Word";
%type "GRAPHIC";  maintain-aspect-ratio TRUE;  display "USEDEF";
%valid_file "F";  width 3.0649in;  height 2.3073in;  depth 0pt;
%original-width 7.2627in;  original-height 5.4518in;  cropleft "0";
%filename 'FIG-arXiv-QET-REVIEW/EPS/fig5.eps';file-properties "XNPEU";}}}%
%BeginExpansion
\begin{center}
\includegraphics[
height=2.3073in,
width=3.0649in
]%
{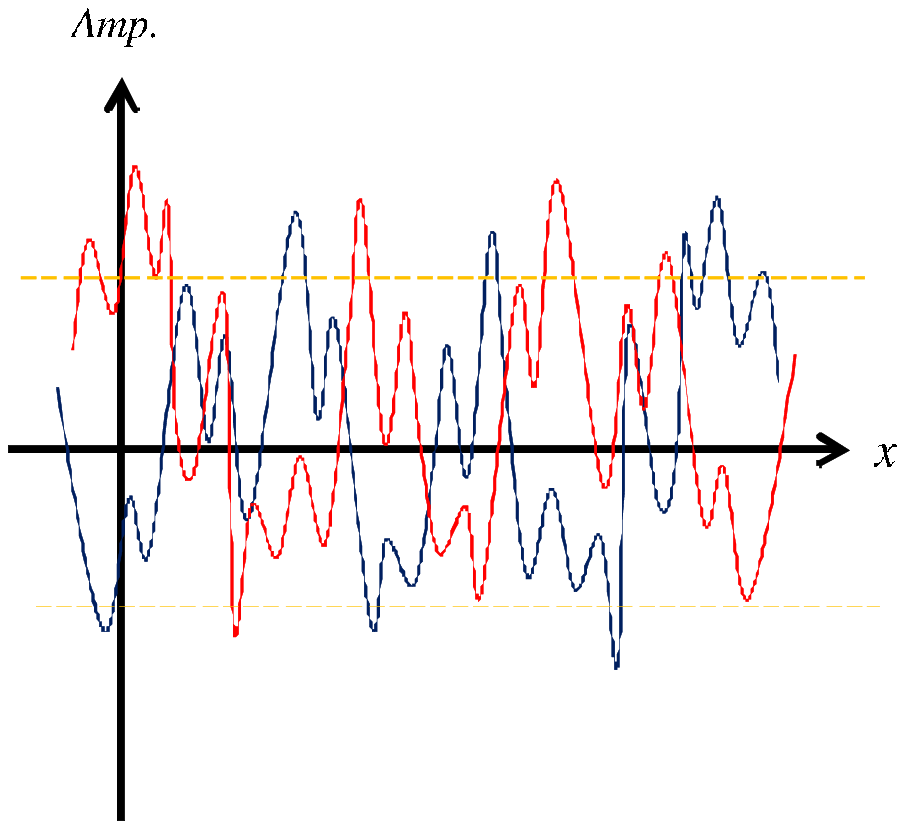}%
\\
{\protect\tiny Figure 5: Zero-point fluctuation of a boson field in one
dimension is schematically depicted. The vertical line implies amplitude of
(coarse-grained) field fluctuation. The horizontal line describes spatial
coordinate x. Mathematical description of the quantum fluctuation is a
superposition of various configuration states. This situation is simplified in
the figure and only two different configurations of the fluctuation are
exhibited by red and blue broken lines. They fluctuate at a typical amplitude
order fixed by the quantum uncertainty relation.}%
\end{center}
%EndExpansion
Of course, quantum fields also have zero-point fluctuation in the vacuum
state. In this case, the fluctuation is also called vacuum fluctuation. In
figure 5, zero-point fluctuation of a boson field in one dimension is
schematically depicted. The vertical line implies amplitude of
(coarse-grained) field fluctuation. The horizontal line describes the spatial
coordinate $x$. The mathematical description of the quantum fluctuation is a
superposition of various configuration states. This situation is simplified in
figure 5 and only two different configurations of the fluctuation are
exhibited by the red and blue broken lines. They are fluctuating with a
typical amplitude order fixed by the quantum uncertainty relation. In later
discussions, let us shift the origin of energy by subtracting the zero-point
energy $E_{g}$ from original energy values so as to make the value of the
ground state zero:%

\[
E^{\prime}=E-E_{g}.
\]
Such a shift is always allowed without changing physics as long as we do not
take into account general relativity. This is because it is not the absolute
value but the difference in energies of two states that is physically
observable. Therefore, the total energy takes nonnegative values. Regardless
of this nonnegativity, quantum theory has a very amazing feature that energy
density can take negative values \cite{bd}. By superposing total-energy
eigenstates, quantum fluctuation in a local region can be more suppressed
(squeezed) than that in the vacuum state via a quantum interference effect.
%TCIMACRO{\FRAME{dtbpFU}{3.0649in}{2.3073in}{0pt}{\Qcb{\QTR{tiny}{Figure 6:
%Illustration of emergence of negative energy density. A typical situation of
%local squeezing of the fluctuation is schematically depicted. The part
%surrounded by an ellipse shows the region of suppressed fluctuation with
%average energy density smaller than that of the vacuum state.}}}{}%
%{fig6.eps}{\special{ language "Scientific Word";  type "GRAPHIC";
%width 3.0649in;  height 2.3073in;  depth 0pt;  original-width 7.2627in;
%original-height 5.4518in;  cropleft "0";  croptop "1";  cropright "1";
%filename 'FIG-arXiv-QET-REVIEW/EPS/fig6.eps';file-properties "XNPEU";}}}%
%BeginExpansion
\begin{center}
\includegraphics[
height=2.3073in,
width=3.0649in
]%
{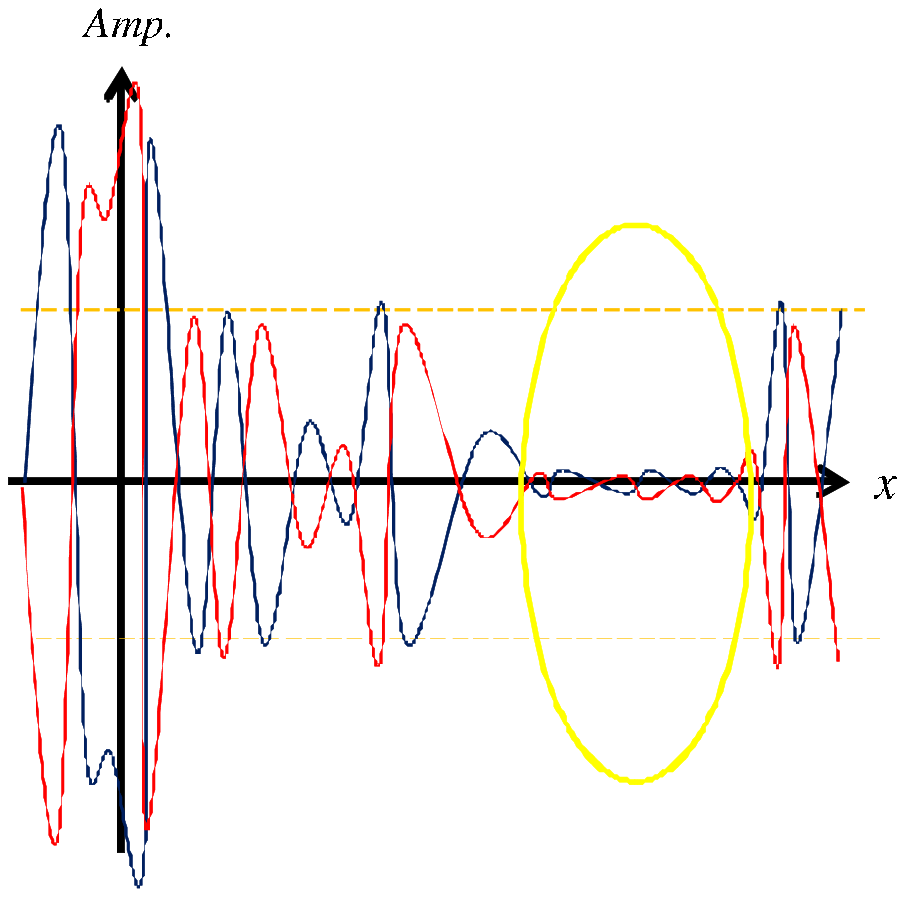}%
\\
{\protect\tiny Figure 6: Illustration of emergence of negative energy density.
A typical situation of local squeezing of the fluctuation is schematically
depicted. The part surrounded by an ellipse shows the region of suppressed
fluctuation with average energy density smaller than that of the vacuum
state.}%
\end{center}
%EndExpansion
In figure 6, a typical situation of local squeezing of the fluctuation is
schematically depicted. The part surrounded by an ellipse shows the region of
suppressed fluctuation with average energy density lower than that of the
vacuum state. As seen in figure 7, energy density in this region must take a
negative value because energy density of the vacuum state is zero and larger
than that of the surrounding region. It is worth stressing here that the total
energy cannot be negative even though we have a region with negative energy
density. This implies that we have other regions with sufficient positive
energy to compensate for this negative energy, as in figure 7.%
%TCIMACRO{\FRAME{dtbpFU}{3.0649in}{2.3073in}{0pt}{\Qcb{\QTR{tiny}{Figure7:
%Energy density distribution in the case of figure 6. }\QTR{normalsize}{
%\ \ \ \ \ \ \ \ \ \ \ \ \ \ \ \ \ \ \ \ }}}{}{fig7.eps}%
%width 3.0649in;  height 2.3073in;  depth 0pt;  original-width 7.2627in;
%original-height 5.4518in;  cropleft "0";  croptop "1";  cropright "1";
%filename 'FIG-arXiv-QET-REVIEW/EPS/fig7.eps';file-properties "XNPEU";}}}%
%BeginExpansion
\begin{center}
\includegraphics[
height=2.3073in,
width=3.0649in
]%
{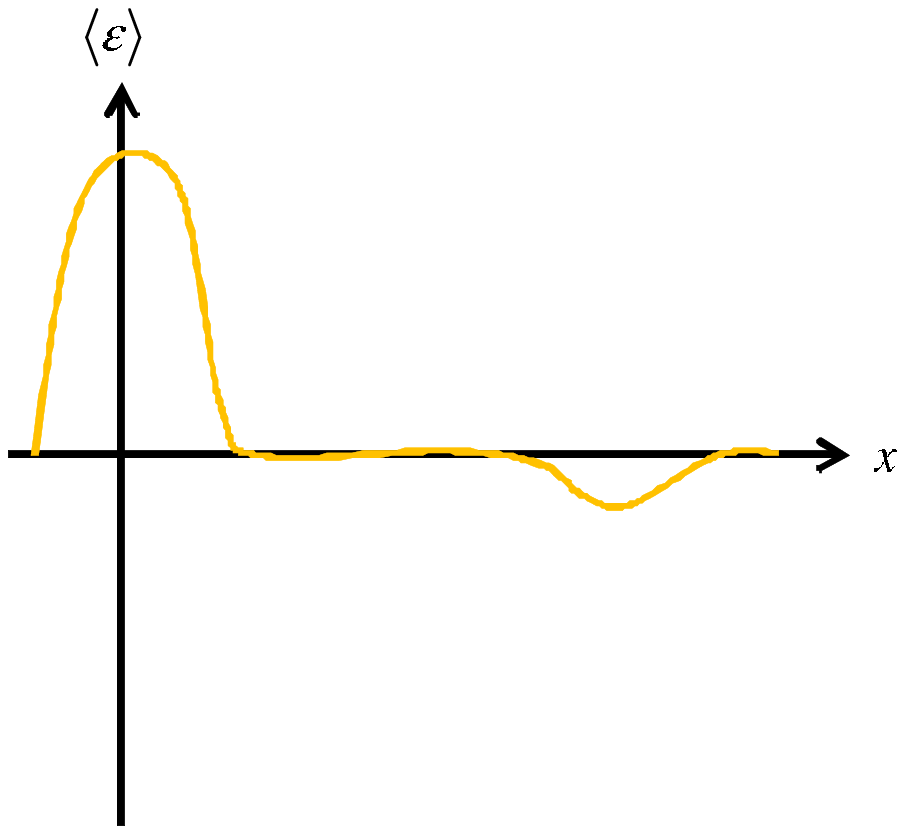}%
\\
{\protect\tiny Figure7: Energy density distribution in the case of figure 6.
}{\protect\normalsize  \ \ \ \ \ \ \ \ \ \ \ \ \ \ \ \ \ \ \ \ }%
\end{center}
%EndExpansion
Local energy, which is an integral of energy density with an appropriate
window function in a compact support, can also take a negative value, i.e., a
value smaller than that of the vacuum state (zero). This fact may tempt us to
directly extract zero-point energy from the vacuum, which really carries zero
local energy \textit{larger than the negative one}. If this was possible, we
would get energy without any cost, but unfortunately, it is not. If we could
extract energy from the vacuum state, the field would be in a state with total
energy less than that in the vacuum state, that is, a negative total-energy
state. However, the total energy must be nonnegative. Therefore, such energy
extraction cannot be attained in physics. For example, if any local unitary
operation $U_{local}$ ($\neq I$) is performed in the vacuum state
$|vac\rangle$, the energy of the field does not decrease but instead
increases. This is because $U_{local}|vac\rangle$ is not the vacuum state but
an excited state \cite{passivity}.%
%TCIMACRO{\FRAME{dtbpFU}{3.0649in}{2.3073in}{0pt}{\Qcb{\QTR{tiny}{Figure 8:
%Energy density distribution after a local unitary operation performed on the
%vacuum state. The energy of the field does not decrease but instead increases
%on average. This is because the state is not the vacuum state but an excited
%state. Of course, the total energy is positive on average. This property is
%called passivity of the vacuum state. }}}{}{fig8.eps}%
%width 3.0649in;  height 2.3073in;  depth 0pt;  original-width 7.2627in;
%original-height 5.4518in;  cropleft "0";  croptop "1";  cropright "1";
%filename 'FIG-arXiv-QET-REVIEW/EPS/fig8.eps';file-properties "XNPEU";}}}%
%BeginExpansion
\begin{center}
\includegraphics[
height=2.3073in,
width=3.0649in
]%
{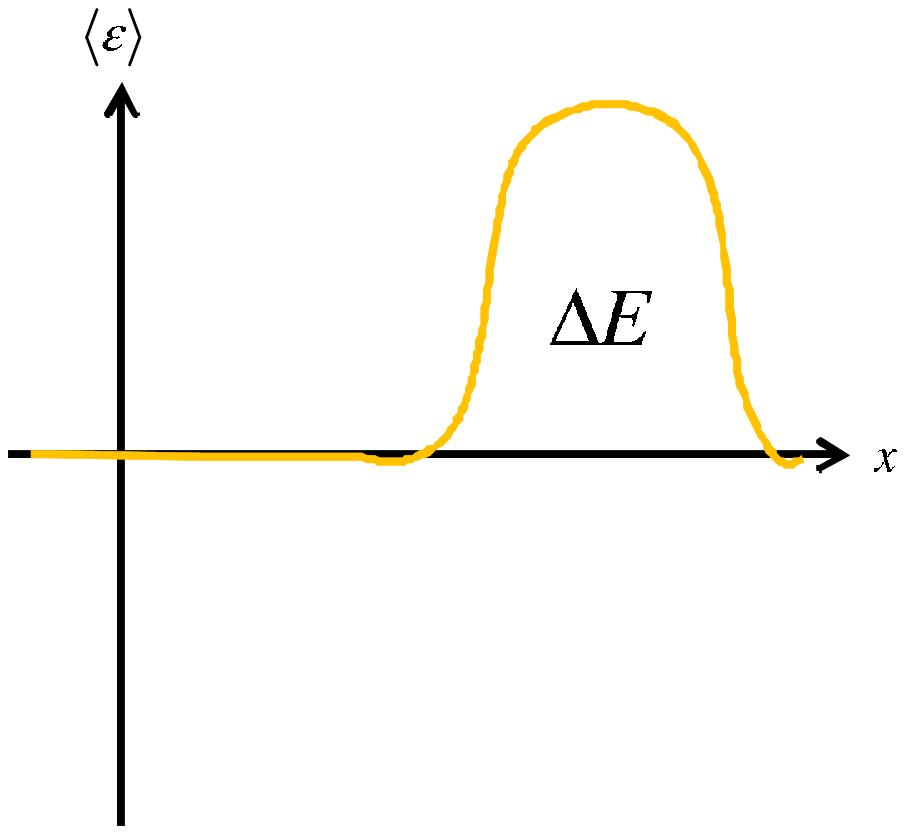}%
\\
{\protect\tiny Figure 8: Energy density distribution after a local unitary
operation performed on the vacuum state. The energy of the field does not
decrease but instead increases on average. This is because the state is not
the vacuum state but an excited state. Of course, the total energy is positive
on average. This property is called passivity of the vacuum state. }%
\end{center}
%EndExpansion
Thus, just like in figure 8, the expectation value of total energy must be
positive:
\begin{equation}
\Delta E=\langle vac|U_{local}^{\dag}HU_{local}|vac\rangle>0.\label{e3}%
\end{equation}
Therefore, the operation requires infusion of the additional energy $\Delta E$
to the field.
%TCIMACRO{\FRAME{dtbpFU}{3.0649in}{2.3073in}{0pt}{\Qcb{\QTR{tiny}{Figure 9:
%Retaining the passivity, the local operation can decrease the amplitude of the
%blue component even though it increases the red component.
%\ \ \ \ \ \ \ \ \ \ \ \ \ \ \ \ \ \ \ \ \ \ \ \ \ \ \ \ \ \ \ \ \ \ \ \ \ \ \ \ \ \ \ \ \ \ \ \ \ \ \ \ \ \ \ \ \ \ \ \ \ \ \ \ \ \ \ \ \ \ \ \ \ \ \ \ \ \ \ \ \ \ \ \ \ \ \ \ \ \ \ \ \ \ \ \ \ \ \ \ \ \ \ \ \ \ \ \ \ \ \ \ \ \ \ \ \ \ \ \ \ \ \ \ \ \ \ \ \ \ \ \ \ \ \ \ \ \ \ \ \ \ \ \ \ \ \ \ \ \ \ \ \ \ \ \ \ \ \ \ \ \ \ \ \ \ \ }}}%
%{}{fig9.eps}{\special{ language "Scientific Word";  type "GRAPHIC";
%width 3.0649in;  height 2.3073in;  depth 0pt;  original-width 7.2627in;
%original-height 5.4518in;  cropleft "0";  croptop "1";  cropright "1";
%filename 'FIG-arXiv-QET-REVIEW/EPS/fig9.eps';file-properties "XNPEU";}}}%
%BeginExpansion
\begin{center}
\includegraphics[
height=2.3073in,
width=3.0649in
]%
{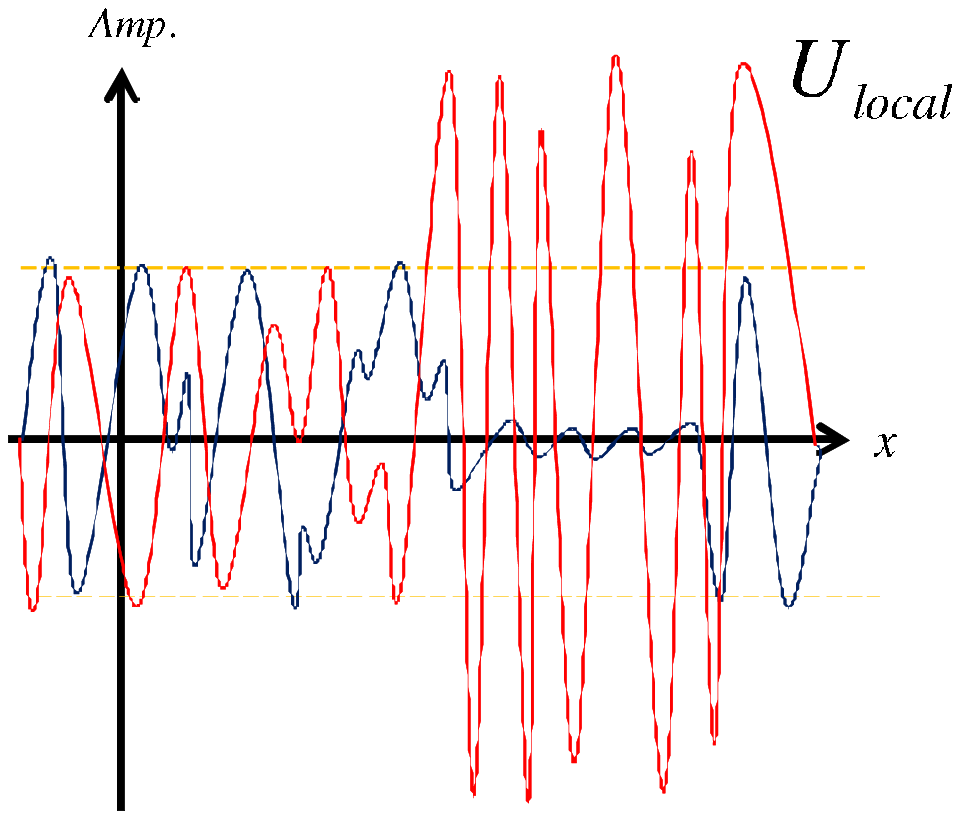}%
\\
{\protect\tiny Figure 9: Retaining the passivity, the local operation can
decrease the amplitude of the blue component even though it increases the red
component.
\ \ \ \ \ \ \ \ \ \ \ \ \ \ \ \ \ \ \ \ \ \ \ \ \ \ \ \ \ \ \ \ \ \ \ \ \ \ \ \ \ \ \ \ \ \ \ \ \ \ \ \ \ \ \ \ \ \ \ \ \ \ \ \ \ \ \ \ \ \ \ \ \ \ \ \ \ \ \ \ \ \ \ \ \ \ \ \ \ \ \ \ \ \ \ \ \ \ \ \ \ \ \ \ \ \ \ \ \ \ \ \ \ \ \ \ \ \ \ \ \ \ \ \ \ \ \ \ \ \ \ \ \ \ \ \ \ \ \ \ \ \ \ \ \ \ \ \ \ \ \ \ \ \ \ \ \ \ \ \ \ \ \ \ \ \ \ }%
\end{center}
%EndExpansion
Here, it should be emphasized that this energy increase takes place just
\textit{on} \textit{average}. It can happen that an operation $U_{local}$
decreases the amplitudes and energy contributions of a few components from
among a large number of superposed fluctuation patterns (the blue component,
for instance, in figure 9). This aspect becomes one of the key points in the
construction of QET later. However, if it happens, other components, like the
red one in figure 9, must be enhanced by much more in their energy
contributions to satisfy the average-value relation in Eq. (\ref{e3}). The
fundamental property in Eq. (\ref{e3}) is called passivity of the vacuum
state. Due to the passivity, one might think that the zero-point energy of the
vacuum state is actually inaccessible free energy hidden in a safe
underground. It really exists in \textit{nothing}, but cannot be harnessed as
long as available operations are local.

Though zero-point energy is totally useless for a single experimenter at a
fixed position, it becomes available if two separate experimenters are able to
perform both local operations and classical communication---this is QET. In
the ground state of an ordinary many-body system, like for a quantum field,
there exists a quantum correlation called \textit{entanglement} \cite{nc}
among zero-point fluctuations of the subsystems. Zero-point fluctuations of
the vacuum in regions A and B are correlated due to the kinetic term of its
Hamiltonian \cite{r}. By virtue of the existence of entanglement, when local
zero-point fluctuation is measured at a position, the measurement result
includes information about quantum fluctuation at a distant position.
%TCIMACRO{\FRAME{dtbpFU}{3.0649in}{2.3073in}{0pt}{\Qcb{\QTR{tiny}{Figure 10:
%QET protocol. At first, zero-point fluctuation is measured in region A and a
%measurement result corresponding to the blue component is obtained in this
%single-shot measurement. The zero-point fluctuation is locally enhanced by
%inputting energy. This result includes information about post-measurement
%quantum fluctuation in region B. The red component vanishes because of the
%wavefunction collapse.  \ \ \ \ \ \ \ \ \ \ \ \ \ \ \ \ \ }}}{}{fig10.eps}%
%width 3.0649in;  height 2.3073in;  depth 0pt;  original-width 7.2627in;
%original-height 5.4518in;  cropleft "0";  croptop "1";  cropright "1";
%filename 'FIG-arXiv-QET-REVIEW/EPS/fig10.eps';file-properties "XNPEU";}}}%
%BeginExpansion
\begin{center}
\includegraphics[
height=2.3073in,
width=3.0649in
]%
{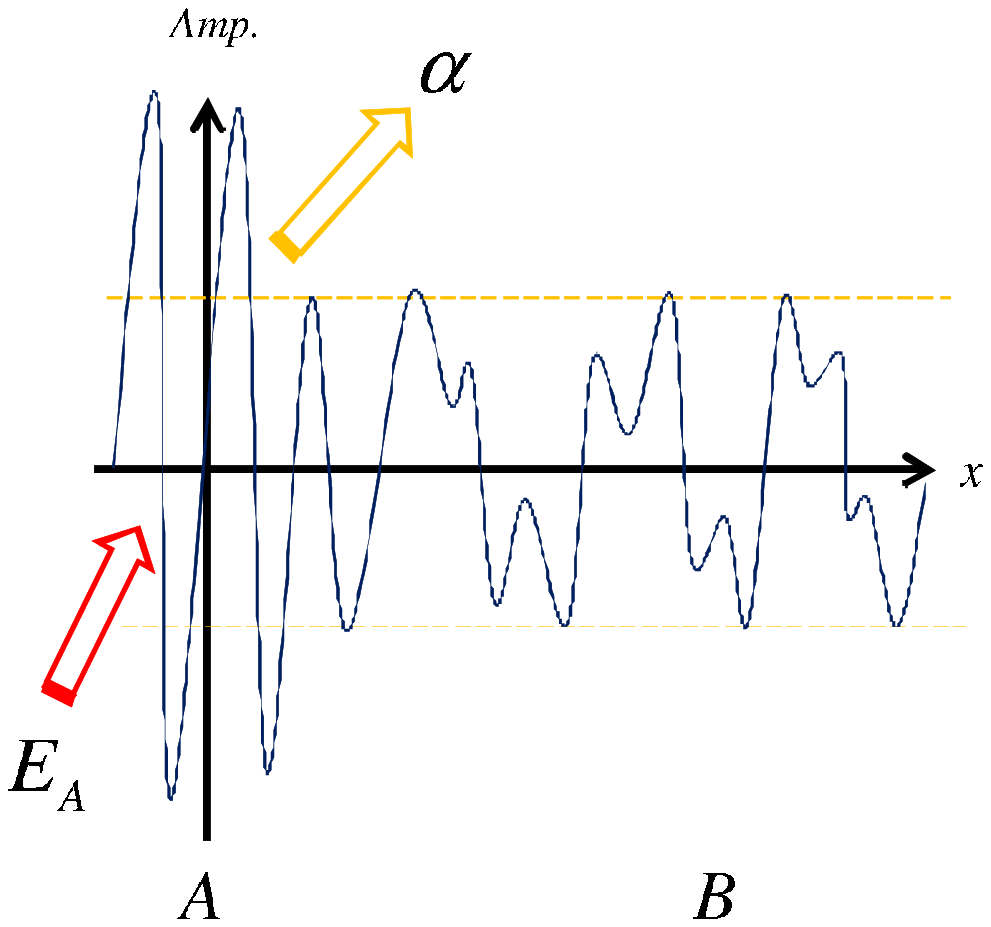}%
\\
{\protect\tiny Figure 10: QET protocol. At first, zero-point fluctuation is
measured in region A and a measurement result corresponding to the blue
component is obtained in this single-shot measurement. The zero-point
fluctuation is locally enhanced by inputting energy. This result includes
information about post-measurement quantum fluctuation in region B. The red
component vanishes because of the wavefunction collapse.
\ \ \ \ \ \ \ \ \ \ \ \ \ \ \ \ \ }%
\end{center}
%EndExpansion
This vacuum-state entanglement is at the heart of the QET protocol with
quantum fields. As the first step of the protocol, zero-point fluctuation is
measured in region A to afford the result $\alpha$ (figure 10). This result
$\alpha$ includes information about post-measurement quantum fluctuation in
region B via entanglement. Hence, we can estimate the quantum fluctuation at B
on the basis of $\alpha$. In the example shown in figure 10, the value of
$\alpha$ corresponding to the blue-line component is obtained by this one-shot
measurement. In this case, the other (red-line) component vanishes because of
the wavefunction collapse when a quantum measurement is performed. (Actually,
practical measurements of local quantum fluctuation are unable to select out a
single configuration of fluctuation, as depicted in figure 10. However, it is
still true that the measurement results include some information about
fluctuation at a distant point, even though the amount of information reduces
as the distance increases.) It should be noted that the measurement device
infuses positive energy $E_{A}$ into the field during this measurement process
because of the vacuum-state passivity, as depicted in figure 11.%
%TCIMACRO{\FRAME{dtbpFU}{3.0649in}{2.3073in}{0pt}{\Qcb{\QTR{tiny}{Figure 11:
%Energy density distribution after the first step of the protocol.}}}%
%{}{fig11.eps}{\special{ language "Scientific Word";  type "GRAPHIC";
%width 3.0649in;  height 2.3073in;  depth 0pt;  original-width 7.2627in;
%original-height 5.4518in;  cropleft "0";  croptop "1";  cropright "1";
%filename 'FIG-arXiv-QET-REVIEW/EPS/fig11.eps';file-properties "XNPEU";}}}%
%BeginExpansion
\begin{center}
\includegraphics[
height=2.3073in,
width=3.0649in
]%
{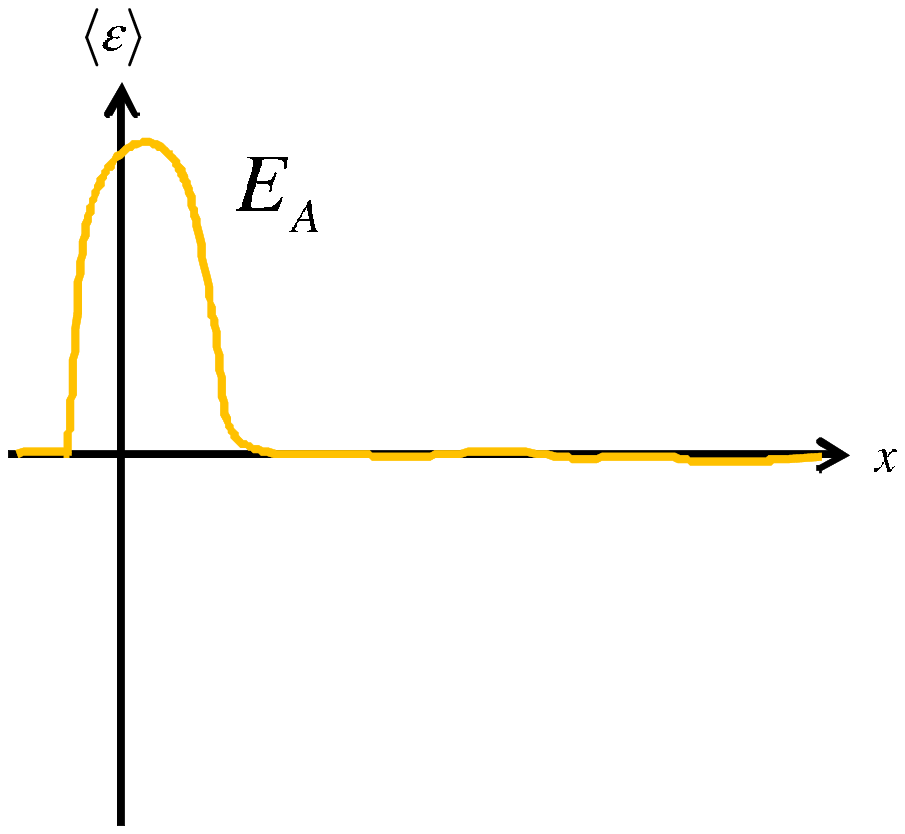}%
\\
{\protect\tiny Figure 11: Energy density distribution after the first step of
the protocol.}%
\end{center}
%EndExpansion
This infused energy is regarded as energy input in the QET protocol. As the
second step, the measurement result $\alpha$ is announced from A to B via a
classical channel. The speed of this announcement can attain the velocity of
light, in principle. During this classical communication, we can neglect the
time evolution of the system, as explained later. On the basis of the
announced $\alpha$, we can devise a strategy, that is, a local unitary
operation $U_{B}(\alpha)$ dependent on $\alpha$, to suppress the realized
quantum fluctuation at B for each value of $\alpha$. As the final step,
$U_{B}(\alpha)$ is performed on the quantum fluctuation of B. This operation
yields negative energy density around B (figure 12) by suppressing only the
amplitude of one component of fluctuation observed in the measurement.%
%TCIMACRO{\FRAME{dtbpFU}{3.0649in}{2.3073in}{0pt}{\Qcb{\QTR{tiny}{Figure 12:
%Energy density distribution after the final step of the protocol.}}}%
%{}{fig12.eps}{\special{ language "Scientific Word";  type "GRAPHIC";
%width 3.0649in;  height 2.3073in;  depth 0pt;  original-width 7.2627in;
%original-height 5.4518in;  cropleft "0";  croptop "1";  cropright "1";
%filename 'FIG-arXiv-QET-REVIEW/EPS/fig12.eps';file-properties "XNPEU";}}}%
%BeginExpansion
\begin{center}
\includegraphics[
height=2.3073in,
width=3.0649in
]%
{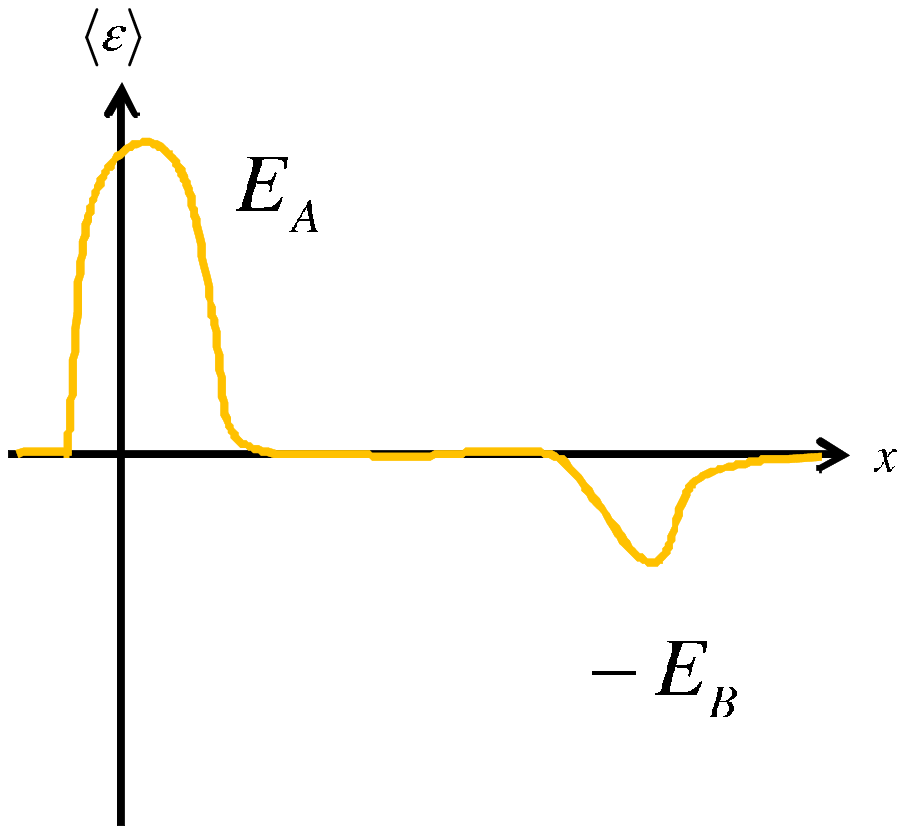}%
\\
{\protect\tiny Figure 12: Energy density distribution after the final step of
the protocol.}%
\end{center}
%EndExpansion%
%TCIMACRO{\FRAME{dtbpFU}{3.0649in}{2.3073in}{0pt}{\Qcb{\QTR{tiny}{Figure 13: At
%the last step of the QET protocol, a unitary operation is performed on quantum
%fluctuation of B. The blue component is suppressed by the operation and yields
%negative local energy. The surplus energy is released as the energy output of
%QET. }}}{}{fig13.eps}{\special{ language "Scientific Word";  type "GRAPHIC";
%width 3.0649in;  height 2.3073in;  depth 0pt;  original-width 7.2627in;
%original-height 5.4518in;  cropleft "0";  croptop "1";  cropright "1";
%filename 'FIG-arXiv-QET-REVIEW/EPS/fig13.eps';file-properties "XNPEU";}}}%
%BeginExpansion
\begin{center}
\includegraphics[
height=2.3073in,
width=3.0649in
]%
{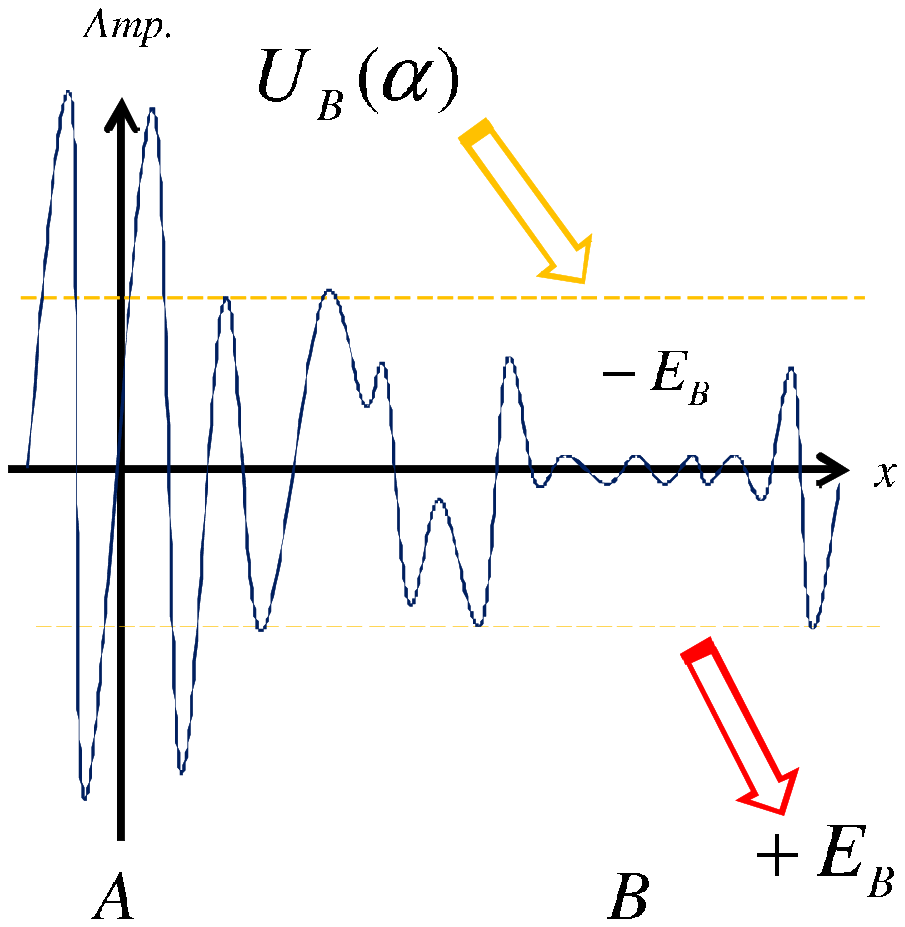}%
\\
{\protect\tiny Figure 13: At the last step of the QET protocol, a unitary
operation is performed on quantum fluctuation of B. The blue component is
suppressed by the operation and yields negative local energy. The surplus
energy is released as the energy output of QET. }%
\end{center}
%EndExpansion
In figure 13, the blue component is suppressed by $U_{B}(\alpha)$. The
operation $U_{B}(\alpha)$ with the value of $\alpha$ corresponding to the blue
component does not need to simultaneously suppress the red component in figure
5 because it has been already been eliminated by this one-shot measurement.
This breaks the passivity barrier against harnessing zero-point energy. After
the last step, the local energy of the field around B takes a negative value
$-E_{B}$. According to local energy conservation, positive energy $+E_{B}$ is
moved from the field to external systems, including the device executing
$U_{B}(\alpha)$. This is regarded as energy output in the QET protocol and can
be harnessed for an arbitrary purpose. Thus, QET really succeeds in effective
energy teleportation in an operational sense. After completion of the
protocol, the total energy of the field is equal to $E_{A}-E_{B}$. Therefore,
the input energy $E_{A}$ is not smaller than the output energy $E_{B}$:%
\[
E_{A}\geq E_{B},
\]
because the total energy does not become negative. Note that the positive
local energy $+E_{A}$ of the field in region A compensates for the negative
local energy $-E_{B}$ of the field in region B in late-time evolution. Hence,
the late-time evolution with cool-down of the system after one round of a
short-acting QET protocol plays a clearing role at the resuming step of the
protocol to prepare for the next round.

This QET mechanism can be summarized using an analogy as follows. The
zero-point energy, which will become the output energy $+E_{B}$ of QET, is
analogous to the energy $E_{zero-point}$ saved in the locked safe underground.
In QET, we get information about a key to the safe, allowing us to extract the
zero-point energy by a remote measurement at A via the vacuum-state
entanglement. However, we must pay the quantum fluctuation at A for this
extraction. The cost is energy $E_{A}$, which is larger than the extracted
zero-point energy $E_{zero-point}(=E_{B})$ taken from the safe at B.

From an operational point of view, the quantum field system can be described
as a microscopic 'energy transporter' from Alice to Bob---yes, like the one in
Star Trek. Before Alice's measurement, the field is in the vacuum state
$\rho_{vac}=|vac\rangle\langle vac|$ carrying no available energy. The
zero-point energy of quantum fluctuation has not yet been activated for use.
By performing a measurement with energy input to Alice's transporter device,
one component of the energetic fluctuation pattern is realized instantaneously
inside Bob's device. More precisely speaking, the post-measurement state
$\rho_{\alpha}$ of the field, which corresponds to the measurement result
$\alpha$, carries the available zero-point field energy, which can be
extracted later by Bob's operation $U_{B}(\alpha)$ generating negative energy
of fluctuation. Note that the energy infused to Alice's device becomes
inactive due to the decrease of the vacuum entanglement during her
measurement, as will be explained in section 4. Using both this protocol of
energy transportation and the standard teleportation protocol for quantum
information, it is possible, in principle, to teleport an object with energy
to a zero-energy local-vacuum region.

Now, the trick to the QET magic considered first is clear. The trick inside A
in figure 2 is the measurement of zero-point fluctuation. The input energy
$E_{A}$ is consumed when performing the measurement. The abracadabra announced
from Alice to Bob is the measurement result $\alpha$. The process inside B in
figure 4 is the local operation $U_{B}(\alpha)$ dependent on $\alpha$. After
the completion of this magic, positive local energy $+E_{A}$ has been hidden
inside A and negative local energy $-E_{B}$ inside B.

In this section, we have omitted the time evolution of quantum fields during
the QET protocol for two reasons: The first is that the field system can be
nonrelativistic. In condensed matter physics, we have many systems, including
quantum Hall edge current \cite{qh} \cite{YIH} and graphene \cite{gr}, that
are described by effective field theory. The speed of energy carriers in these
systems is much smaller than the velocity of light. Therefore, the local
operations and classical communication of QET can be assumed to take a very
short time, during which we may neglect time evolution of the effective
fields. The second reason is that a QET process in which time evolution is
essentially irrelevant can be actually constructed using relativistic fields
\cite{8}. We are able to consider a setting in which wave packets excited by
Alice's measurement do not propagate toward Bob in time evolution. Thus, Alice
can send only information to Bob, avoiding directly sending Bob the energy
emitted from the measurement device. This allows us to establish a nontrivial
QET protocol between them. In fact, such a QET protocol with a relativistic
field is introduced in section 5.

Before closing this section, it is worth stressing that we do not need to
worry about time-energy uncertainty relations for measurements of the output
energy $E_{B}$ of QET. In contrast to the position-momentum uncertainty
relation, time-energy uncertainty relations are not fundamental laws. They are
conceptually dependent on purposes of our tasks and models of energy
measurements, and do not have a universal meaning in quantum theory
\cite{teur}. This is essentially because time in quantum theory is not an
observable to be measured but an external parameter. We must fix in advance a
definite time slice in our spacetime to define a quantum state of a system and
perform measurements of arbitrary observables at the time. The Hamiltonian,
which stands for the total energy of the system, is just one of ordinary
observables we can measure instantaneously at a fixed time slice. In fact, the
very famous pointer-basis measurement proposed by von Neumann \cite{vN} is
capable of attaining an instantaneous measurement of energy as follows. Let us
consider a system $S$ with Hamiltomian $H_{S}$ and a probe system to measure
energy of $S$. The probe is a quantum particle in an infinite line
parametrized by a spatial coordinate $x$. The position $x$ of the particle is
interpreted as the position of the energy pointer of the measurement device.
Let us assume that the initial state of the pointer particle is localized at
$x=0$ as
\[
\psi_{P}(x)=\delta(x).
\]
The pointer particle has no free Hamiltonian, but couples with $S$ via a
measurement interaction given by
\begin{equation}
H_{m}(t)=\delta(t)H_{S}\otimes(-i\hbar\partial_{x}). \label{1e}%
\end{equation}
Let us prepare a state of $S$ such that
\[
|\Psi_{S}\rangle=\sum_{n}c_{n}|E_{n}\rangle,
\]
where $c_{n}$ are complex coefficients, and $|E_{n}\rangle$ is an eigenstate
of $H_{S}$ corresponding to an eignvalue $E_{n}$. After the instantaneous
interaction (\ref{1e}), the state of the composite system becomes%
\[
\exp\left[  -H_{S}\otimes\partial_{x}\right]  \left(  |\Psi_{S}\rangle
\otimes\psi_{P}(x)\right)  =\sum_{n}c_{n}|E_{n}\rangle\otimes\delta(x-E_{n}).
\]
Soon after the switch off of the interaction, we can perform a projective
measurement of the position $x$ of the pointer particle. This gives a value of
energy of $S$ at $t=0$ as a single-shot measurement result. Therefore energy
can be measured instantaneously. Meanwhile, if one may consider a bad class of
energy measurements, time-energy uncertainty relations hold and prevent us
from measuring energy precisely. For example, tracking time evolution of a
state of a system during a time $T$ allows us to estimate energy of the system
by using time-Fourier transformation of the state. However, the estimation has
inevitable error of order of $\hbar/T$ in a similar way to the momentum
measurement of a particle confined in a finite spatial region. In our QET
analysis, we do not adopt such bad measurements governed by (non-universal)
time-energy uncertainty relations. In the reference \cite{YIH} and the last
part of the next section, more realistic models of extraction and measurement
of $E_{B}$ are discussed.

\bigskip

\section{Minimal QET Model}

\bigskip

~

In this section, the most simple example of QET is reviewed. We adopt the
natural unit $\hbar=1$. For a detailed analysis, see \cite{3}. The system
consists of two qubits $A$ and $B$. Its Hamiltonian reads
\[
H=H_{A}+H_{B}+V,
\]
where each contribution is given by
\begin{align}
H_{A}  &  =h\sigma_{A}^{z}+\frac{h^{2}}{\sqrt{h^{2}+k^{2}}},\label{e4}\\
H_{B}  &  =h\sigma_{B}^{z}+\frac{h^{2}}{\sqrt{h^{2}+k^{2}}},\\
V  &  =2k\sigma_{A}^{x}\sigma_{B}^{x}+\frac{2k^{2}}{\sqrt{h^{2}+k^{2}}},
\label{e5}%
\end{align}
and $h$ and$~k$ are positive constants with energy dimensions, $\sigma_{A}%
^{x}~\left(  \sigma_{B}^{x}\right)  $ is the $x$-component of the Pauli
operators for the qubit $A$ ($B$), and $\sigma_{A}^{z}~\left(  \sigma_{B}%
^{z}\right)  $ is the $z$-component for the qubit $A$ ($B$). The constant
terms in Eqs. (\ref{e4})--(\ref{e5}) are added in order to make the
expectation value of each operator zero for the ground state $|g\rangle$:
\[
\langle g|H_{A}|g\rangle=\langle g|H_{B}|g\rangle=\langle g|V|g\rangle=0.
\]
Because the lowest eigenvalue of the total Hamiltonian $H$ is zero, $H$ is a
nonnegative operator:$~$%
\[
H\geq0,
\]
which means that expectation value of $H$ for an arbitrary state $|\Psi
\rangle$ is nonnegative:%

\[
\langle\Psi|H|\Psi\rangle\geq0.
\]
Meanwhile, it should be noted that $H_{B}$ and $H_{B}+V$ have negative
eigenvalues, which can yield negative energy density at $B$. The ground state
is given by
\begin{align*}
|g\rangle &  =\frac{1}{\sqrt{2}}\sqrt{1-\frac{h}{\sqrt{h^{2}+k^{2}}}}%
|+\rangle_{A}|+\rangle_{B}\\
&  -\frac{1}{\sqrt{2}}\sqrt{1+\frac{h}{\sqrt{h^{2}+k^{2}}}}|-\rangle
_{A}|-\rangle_{B},
\end{align*}
where $|\pm\rangle_{A}~\left(  |\pm\rangle_{B}\right)  $ is the eigenstate of
$\sigma_{A}^{z}~\left(  \sigma_{B}^{z}\right)  $ with eigenvalue $\pm1$. A QET
protocol is constructed by the following three steps:

\begin{itemize}
\item I. A projective measurement of observable $\sigma_{A}^{x}$ is performed
on $A$ in the ground state $|g\rangle$ and a measurement result $\alpha=\pm1$
is obtained. During the measurement, a positive amount of energy%
\begin{equation}
E_{A}=\frac{h^{2}}{\sqrt{h^{2}+k^{2}}} \label{e6}%
\end{equation}
is infused to $A$ on average.

\item II. The result $\alpha$ is announced to $B$ via a classical channel at a
speed much faster than the velocity of energy diffusion of the system.

\item III. Let us consider a local unitary operation on $B$ depending on the
value of $\alpha$ given by
\[
U_{B}(\alpha)=I_{B}\cos\theta-i\alpha\sigma_{B}^{y}\sin\theta,
\]
where $\theta$ is a real constant that satisfies%
\begin{align}
\cos\left(  2\theta\right)   &  =\frac{h^{2}+2k^{2}}{\sqrt{\left(
h^{2}+2k^{2}\right)  ^{2}+h^{2}k^{2}}},\label{e12}\\
\sin\left(  2\theta\right)   &  =\frac{hk}{\sqrt{\left(  h^{2}+2k^{2}\right)
^{2}+h^{2}k^{2}}}. \label{e13}%
\end{align}
$U_{B}(\alpha)$ is performed on $B$. During the operation, a positive amount
of energy
\begin{equation}
E_{B}=\frac{h^{2}+2k^{2}}{\sqrt{h^{2}+k^{2}}}\left[  \sqrt{1+\frac{h^{2}k^{2}%
}{\left(  h^{2}+2k^{2}\right)  ^{2}}}-1\right]  \label{e7}%
\end{equation}
$~$ is extracted from $B$ on average.

Firstly, the projection operator corresponding to each measurement result
$\alpha$ of $\sigma_{A}^{x}$ is given by
\[
P_{A}(\alpha)=\frac{1}{2}\left(  1+\alpha\sigma_{A}^{x}\right)  .
\]
The post-measurement state of the two qubits with output $\alpha$ is given by
\[
|A(\alpha)\rangle=\frac{1}{\sqrt{p_{A}(\alpha)}}P_{A}(\alpha)|g\rangle,
\]
where $p_{A}(\alpha)$ is the emergence probability of $\alpha$ for the ground
state and given by $\langle g|P_{A}(\alpha)|g\rangle$. It is easy to check
that the average post-measurement state given by
\[
\sum_{\alpha}p_{A}(\alpha)|A(\alpha)\rangle\langle A(\alpha)|=\sum_{\alpha
}P_{A}(\alpha)|g\rangle\langle g|P_{A}(\alpha)
\]
has a positive expectation value $E_{A}$ of $H$, which has an energy
distribution localized at $A$. In fact, the value defined by
\begin{equation}
E_{A}=\sum_{\alpha}\langle g|P_{A}(\alpha)HP_{A}(\alpha)|g\rangle=\sum
_{\alpha}\langle g|P_{A}(\alpha)H_{A}P_{A}(\alpha)|g\rangle\label{e9}%
\end{equation}
can be computed straightforwardly and Eq.(\ref{e6}) is obtained. This infused
energy $E_{A}$ is regarded as the QET energy input via the measurement of $A$.
During the measurement, $E_{A}$ is transferred from external systems,
including the measurement device and its power source, respecting local energy conservation.
\end{itemize}

The nontrivial feature of this measurement is that it does not increase the
average energy of $B$ at all. By explicit calculations using
\[
\left[  \sigma_{A}^{x},H_{B}\right]  =\left[  \sigma_{A}^{x},V\right]  =0,
\]
the average values of $H_{B}$ and $V$ are found to remain zero after the
measurement and are the same as those in the ground state:%
\begin{align*}
\sum_{\alpha}\langle g|P_{A}(\alpha)H_{B}P_{A}(\alpha)|g\rangle &  =\langle
g|H_{B}|g\rangle=0,\\
\sum_{\alpha}\langle g|P_{A}(\alpha)VP_{A}(\alpha)|g\rangle &  =\langle
g|V|g\rangle=0.
\end{align*}
Thus we cannot extract energy from $B$ by local operations
\textit{independent} of $\alpha$ soon after the measurement. To verify this
fact explicitly, let us consider a local unitary operation $W_{B}$ independent
of $\alpha$ and performed on $B$.\quad Then the post-operation state $\omega$
is given by
\begin{align*}
\omega &  =\sum_{\alpha}W_{B}P_{A}(\alpha)|g\rangle\langle g|P_{A}%
(\alpha)W_{B}^{\dag}\\
&  =W_{B}\left(  \sum_{\alpha}P_{A}(\alpha)|g\rangle\langle g|P_{A}%
(\alpha)\right)  W_{B}^{\dag}.
\end{align*}
The energy difference after the operation is calculated as
\begin{equation}
E_{A}-\operatorname*{Tr}\left[  \omega H\right]  =-\langle g|W_{B}^{\dag
}\left(  H_{B}+V\right)  W_{B}|g\rangle, \label{e8}%
\end{equation}
where we have used
\[
W_{B}^{\dag}H_{A}W_{B}=H_{A}W_{B}^{\dag}W_{B}=H_{A},
\]%
\[
\left[  W_{B}^{\dag}\left(  H_{B}+V\right)  W_{B},~P_{A}(\alpha)\right]  =0,
\]
and the completeness relation of $P_{A}(\alpha)$:%
\[
\sum_{\alpha}P_{A}(\alpha)=1_{A}.
\]
From Eq. (\ref{e8}), it is proven that the energy difference is not positive:
\[
E_{A}-\operatorname*{Tr}\left[  \omega H\right]  =-\langle g|W_{B}^{\dag
}HW_{B}|g\rangle\leq0,
\]
because of a relation such that
\[
\langle g|W_{B}^{\dag}H_{A}W_{B}|g\rangle=\langle g|H_{A}|g\rangle=0
\]
and the nonnegativity of $H$. Therefore, as a natural result, no local
operation on $B$ independent of $\alpha$ extracts positive energy from $B$ by
decreasing the total energy of the two qubits.

The infused energy $E_{A}$ diffuses to $B$ after a while. The time evolution
of the expectation values $H_{B}$ and $V$ of the average post-measurement
state is calculated as%
\begin{align*}
\langle H_{B}(t)\rangle &  =\sum_{\alpha}\langle g|P_{A}(\alpha)|g\rangle
\langle A(\alpha)|e^{itH}H_{B}e^{-itH}|A(\alpha)\rangle\\
&  =\frac{h^{2}}{2\sqrt{h^{2}+k^{2}}}\left[  1-\cos\left(  4kt\right)
\right]  ,
\end{align*}
and
\[
\langle V(t)\rangle=\sum_{\alpha}\langle g|P_{A}(\alpha)|g\rangle\langle
A(\alpha)|e^{itH}Ve^{-itH}|A(\alpha)\rangle=0.
\]
Therefore, we enable energy to be extracted from $B$ after a diffusion time
scale of $1/k$; this is just the more pedestrian form of energy transportation
from $A$ to $B$. The QET protocol achieves energy transportation from $A$ to
$B$ in a time scale much shorter than that of this conventional transportation.

In step II of the protocol, the measurement output $\alpha$ is announced to
$B$. Because the model is nonrelativistic, the propagation speed of the
announced output can be much faster than the diffusion speed of the infused
energy and can be approximated as infinity. Soon after the arrival of the
output $\alpha$, $U_{B}(\alpha)$ is performed on $B$. Then the average state
after the operation is given by%

\[
\rho=\sum_{\alpha}U_{B}(\alpha)P_{A}(\alpha)|g\rangle\langle g|P_{A}%
(\alpha)U_{B}(\alpha)^{\dag}.
\]
The expectation value of the total energy after the operation is given by
\[
\operatorname*{Tr}\left[  \rho H\right]  =\sum_{\alpha}\langle g|P_{A}%
(\alpha)U_{B}(\alpha)^{\dag}HU_{B}(\alpha)P_{A}(\alpha)|g\rangle.
\]
On the basis of the fact that $U_{B}(\alpha)$ commutes with $H_{A}$ and Eq.
(\ref{e9}), the output energy $E_{B}$ is computed as
\[
E_{B}=E_{A}-\operatorname*{Tr}\left[  \rho H\right]  =-\operatorname*{Tr}%
\left[  \rho\left(  H_{B}+V\right)  \right]  .
\]
Further, on the basis of the fact that $P_{A}(\alpha)$ commutes with
$U_{B}(\alpha)$, $H_{B}$, and $V$, the energy can be written as
\[
E_{B}=-\sum_{\alpha}\langle g|P_{A}(\alpha)\left(  H_{B}(\alpha)+V(\alpha
)\right)  |g\rangle,
\]
where the $\alpha$-dependent operators are given by
\begin{align*}
H_{B}(\alpha)  &  =U_{B}(\alpha)^{\dag}H_{B}U_{B}(\alpha),\\
V(\alpha)  &  =U_{B}(\alpha)^{\dag}VU_{B}(\alpha).
\end{align*}
By a straightforward calculation, $E_{B}$ is computed as
\begin{equation}
E_{B}=\frac{1}{\sqrt{h^{2}+k^{2}}}\left[  hk\sin(2\theta)-\left(  h^{2}%
+2k^{2}\right)  \left[  1-\cos\left(  2\theta\right)  \right]  \right]  .
\label{e10}%
\end{equation}
Note that $E_{B}=0$ if $\theta=0$, as it should be. If we take a small
positive value of $\theta$ in Eq. (\ref{e10}), note that $E_{B}$ takes a small
positive value such that%
\[
E_{B}\sim\frac{2hk\left\vert \theta\right\vert }{\sqrt{h^{2}+k^{2}}}>0.
\]
Maximization of $E_{B}$ in terms of $\theta$ is achieved by taking a value of
$\theta$ that satisfies Eqs. (\ref{e12}) and (\ref{e13}). Substituting Eqs.
(\ref{e12}) and (\ref{e13}) into Eq. (\ref{e10}) yields the positive value of
$E_{B}$ in Eq. (\ref{e7}). Therefore, even though energy carriers coming from
$A$ have not yet arrived at $B$, the QET protocol can achieve energy
extraction from $B$. As stressed in section 2, the success of energy
extraction is due to the emergence of negative energy density at $B$. In the
section of summary and comment, it will be discussed that a large amount of
teleported energy requests a large amount of consumption of the ground-state
entanglement between A and B in this model.

Finally a comment is added about extraction and measurement of $E_{B}$. As
mentioned in the previous section, there is a nontrivial aspect of energy
measurements. Some bad measurements suffer from time-energy uncertainty
relations and give inevitable error in estimation of $E_{B}$. However, we can
avoid such a risk by adopting other good measurements of energy. The
pointer-basis measurement is one of such good measurements, as stressed in the
previous section. Here another setup to measure $E_{B}$ \cite{2} \cite{4} is
reviewed compatible with realistic experiments of QET. After the arrival of
the measurement result $\alpha$ at the region of $B$, let us generate a laser
pulse $W(\alpha)$ in an optical fiber which polarization is dependent on
$\alpha$. A spatial coordinate $\zeta$ parametrizes the fiber. The fiber is
connected between the generation point of $W(\alpha)$ ($\zeta=\zeta_{i}$%
)$~$and the final point $\left(  \zeta=\zeta_{f}\right)  $ via a point
$\zeta=\zeta_{B}$ where spin $B$ stays: $\zeta_{i}<\zeta_{B}<\zeta_{f}$. The
pulse $W(\alpha)$ moves toward the final point, and intersects with spin $B$
at $\zeta=\zeta_{B}$ on the way. Then it can be verified as follows that
$W(\alpha)$ performs $U_{B}(\alpha)$ to $B$. Let us introduce creation and
annihilation bosonic operators $\Psi_{\alpha}^{\dag}(\zeta)$ and $\Psi
_{\alpha}(\zeta)$ for one photon of the laser field with polarization
$\alpha=\pm$ in the fiber. The operators $\Psi_{\alpha}^{\dag}(\zeta)$ and
$\Psi_{\alpha}(\zeta)$ satisfy the following commutation relations:%
\begin{align*}
\left[  \Psi_{\alpha}(\zeta),~\Psi_{\alpha^{\prime}}^{\dag}(\zeta^{\prime
})\right]   &  =\delta_{\alpha\alpha^{\prime}}\delta\left(  \zeta
-\zeta^{\prime}\right)  ,\\
\left[  \Psi_{\alpha}(\zeta),~\Psi_{\alpha^{\prime}}(\zeta^{\prime})\right]
&  =0,\\
\left[  \Psi_{\alpha}^{\dag}(\zeta),~\Psi_{\alpha^{\prime}}^{\dag}%
(\zeta^{\prime})\right]   &  =0.
\end{align*}
The vacuum state $|0\rangle$ of the laser field is defined by%
\[
\Psi_{\alpha}(\zeta)|0\rangle=0.
\]
Let us assume that the initial state of the laser field is a pulse-wave
coherent state with polarization $\alpha$ given by
\[
|\alpha\rangle\propto\exp\left(  \int_{-\infty}^{\infty}f_{i}(\zeta
)\Psi_{\alpha}^{\dag}(\zeta)d\zeta\right)  |0\rangle,
\]
where $f_{i}(\zeta)$ is the coherent amplitude of the state and a function
with a support localized around $\zeta=\zeta_{i}$. The field strength of the
pulse is defined by%
\[
F=\int_{-\infty}^{\infty}\left\vert f_{i}(\zeta)\right\vert ^{2}d\zeta.
\]
In order to consider a semi-classical coherent state, let us take a large
value of $F$. The free Hamiltonian of the fiber photon reads
\[
H_{\Psi}=-\frac{ic}{2}\int_{-\infty}^{\infty}\left[  \Psi(\zeta)^{\dagger
}\partial_{\zeta}\Psi(\zeta)-\partial_{\zeta}\Psi(\zeta)^{\dagger}\Psi
(\zeta)\right]  d\zeta,
\]
where $c$ is the light velocity in the fiber and $\Psi(\zeta)$ is given by
\[
\Psi(\zeta)=\left[
\begin{array}
[c]{c}%
\Psi_{+}(\zeta)\\
\Psi_{-}(\zeta)
\end{array}
\right]  .
\]
The free evolution of the photon field is given by%
\[
e^{itH_{\Psi}}\Psi(\zeta)e^{-itH_{\Psi}}=\Psi(\zeta-ct).
\]
The laser field couples with spin $B$ via the interaction given by%

\begin{equation}
H_{LO}=\frac{c}{dF}\theta\sigma_{B}^{y}\int_{\zeta_{B}-d/2}^{\zeta_{B}%
+d/2}\left[  \Psi_{+}^{\dagger}(\zeta)\Psi_{+}(\zeta)-\Psi_{-}^{\dagger}%
(\zeta)\Psi_{-}(\zeta)\right]  d\zeta, \label{3e}%
\end{equation}
where $d$ is the width of the interaction region. The total Hamiltonian of the
composite system is expressed as%
\[
H_{tot}=H+H_{LO}+H_{\Psi},
\]
and conserved in time. Before the intersection of $W(\alpha)$ with $B$, the
initial state of the composite system is given by%
\[
\frac{P_{A}(\alpha)|g\rangle\langle g|P_{A}(\alpha)}{\langle g|P_{A}%
(\alpha)|g\rangle}\otimes|\alpha\rangle\langle\alpha|.
\]
In this model, the evolution of the laser pulse induces effective switching of
the interaction for $U_{B}(\alpha)$. In fact, the interaction $H_{LO}$ in Eq.
(\ref{3e}) is active only when the pulse exists inside $\left[  \zeta
_{B}-d/2,\zeta_{B}+d/2\right]  $. Because $W(\alpha)$ is a semi-classical
coherent pulse with large $F$, the photon field can be treated as an external
classical field for $B$ in the leading approximation. The switching process
for $B$ is described by an effective interaction Hamiltonian as $\langle
\alpha|H_{LO}(t)|\alpha\rangle$, where $H_{LO}(t)=e^{itH_{\Psi}}%
H_{LO}e^{-itH_{\Psi}}$. Assuming that the width of the pulse form $f_{i}%
(\zeta)$ is much smaller than $d$, $\langle\alpha|H_{LO}(t)|\alpha\rangle$ can
be approximated as
\[
\langle\alpha|H_{LO}(t)|\alpha\rangle\sim\frac{c}{d}\Theta\left(  \frac{d}%
{2c}-\left\vert t\right\vert \right)  \theta\sigma_{B}^{y}\frac{\alpha}{F}%
\int_{-\infty}^{\infty}\left\vert f_{i}(\zeta)\right\vert ^{2}d\zeta\sim
\delta\left(  t\right)  \alpha\theta\sigma_{B}^{y}%
\]
by taking the nonrelativistic limit ($c\sim\infty$). Then the time evolution
operator of $B$ induced by this effective interaction is calculated as%

\[
\operatorname*{T}\exp\left[  -i\int_{-0}^{+0}\langle\alpha|H_{LO}%
(t)|\alpha\rangle dt\right]  =\exp\left[  -i\alpha\theta\sigma_{B}^{y}\right]
=U_{B}(\alpha).
\]
Thus, the interaction in Eq. (\ref{3e}) certainly reproduces the operation
$U_{B}(\alpha)$. The energy of $W(\alpha)$ changes when the pulse passes
through $\left[  \zeta_{B}-d/2,\zeta_{B}+d/2\right]  $. The average energy of
the two-spin system before the interaction with the pulse is $E_{A}$. The
initial averaged energy of the pulse is denoted by $E_{1}$. The average energy
of the two-spin system\ after the interaction becomes $E_{A}-E_{B}$ because
$U_{B}(\alpha)$ is operated to $B$ and the energy decreases as the QET effect.
The averaged pulse energy after the interaction is denoted by $E_{2}$. Then
the conservation of $H_{tot}$ ensures that
\begin{equation}
E_{A}+E_{1}=\left(  E_{A}-E_{B}\right)  +E_{2}, \label{77e}%
\end{equation}
because $H_{LO}$ has no contribution in the initial and final state of the
scattering process between $B$ and the pulse. Using Eq. (\ref{77e}), the
output energy of QET can be rewritten as $E_{B}=E_{2}-E_{1}$. Consequently, by
measuring the initial and final energy of the pulse many times and taking the
averages, we can precisely determine the output energy of QET without any
problems caused by time-energy uncertainty relations.

\bigskip

\section{General Theory of Quantum Energy Teleportation}

~

In this section, the general theory of QET is introduced for one-dimensional
discrete chain models. The model is a system composed of many quantum
subsystems of general types arrayed in one dimension. The subsystems, labeled
by site numbers $n$, are coupled with each other via nearest-neighbor
interactions, as depicted in figure 14.%
%TCIMACRO{\FRAME{dtbpFU}{3.0649in}{2.3073in}{0pt}{\Qcb{\QTR{tiny}{Figure 14:
%Schematical figure of the chain model with nearest-neighbor interaction. The
%circles stand for quantum subsystems. The chain is labeled by site numbers n.
%}}}{}{fig14.eps}{\special{ language "Scientific Word";  type "GRAPHIC";
%width 3.0649in;  height 2.3073in;  depth 0pt;  original-width 7.2627in;
%original-height 5.4518in;  cropleft "0";  croptop "1";  cropright "1";
%filename 'FIG-arXiv-QET-REVIEW/EPS/fig14.eps';file-properties "XNPEU";}}}%
%BeginExpansion
\begin{center}
\includegraphics[
height=2.3073in,
width=3.0649in
]%
{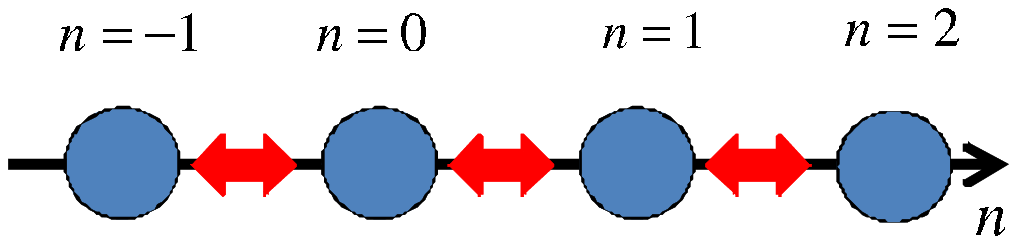}%
\\
{\protect\tiny Figure 14: Schematical figure of the chain model with
nearest-neighbor interaction. The circles stand for quantum subsystems. The
chain is labeled by site numbers n. }%
\end{center}
%EndExpansion
We adopt $\hbar=1$ unit and concentrate on a short time scale during which
dynamical evolution induced by the Hamiltonian $H$ is negligible. Let us
denote the difference between the largest and smallest eigenvalues of $H$ by
$\Delta E$. The timescale $t$ discussed here is assumed to satisfy%
\begin{equation}
t\ll\frac{1}{\Delta E}.\label{sts}%
\end{equation}
Assuming this condition, it is valid to treat the time evolution operator as
$\exp\left[  -itH\right]  \sim I$. It should also be noted that the condition
in Eq. (\ref{sts}) can be weakened in the case that a finite amount of energy
$E_{in}$ less than $\Delta E$ is inputted to the system in the ground state,
as follows:
\[
t\ll\frac{1}{E_{in}}.
\]
In addition, let us assume that LOCC can be repeated for the subsystems many
times even in the short time interval. If the site number difference between
two parties in the protocol is given by $\Delta n$ and the chain spacing
between nearest-neighbor sites is given by $a$, the time scale condition for
many-round LOCC is expressed as%

\begin{equation}
t\gg a\Delta n/c, \label{locc}%
\end{equation}
where $c$ is the velocity of light. By taking the nonrelativistic limit
$c\rightarrow\infty$, the relation in Eq.(\ref{locc}) always holds.

The energy density operators are Hermitian operators and take the general
forms of%

\begin{equation}
T_{n}=X_{n}+\sum_{j}\left(  \frac{1}{2}g_{n-1/2,j}Y_{n-1,j}Y_{n,j}+\frac{1}%
{2}g_{n+1/2,j}Y_{n,j}Y_{n+1,j}\right)  , \label{e11}%
\end{equation}
where $X_{n}$ and $Y_{n,j}$ are local operators for the subsystem at site $n$,
and $g_{n\pm1/2,j}$ are real coupling constants for the nearest-neighbor
interaction. The total Hamiltonian is given by a site-sum of $T_{n}$:
\[
H=\sum_{n}T_{n}.
\]
The ground state $|g\rangle$ is the eigenstate of $H$ with the lowest
eigenvalue. If the expectation values of $T_{n}$ do not vanish for the ground
state as
\[
\langle g|T_{n}|g\rangle=\epsilon_{n}\neq0,
\]
we shift the operator $X_{n}$ on the right-hand side of Eq. (\ref{e11}) by the
constant $\epsilon_{n}$ as
\[
X_{n}^{\prime}=X_{n}-\epsilon_{n},
\]
without changing physics. Then, without loss of generality, we can assume the
relation given by%

\begin{equation}
\langle g|T_{n}|g\rangle=0. \label{12}%
\end{equation}
Eq. (\ref{12}) derives that the eigenvalue of $H$ for the ground state is zero:%

\begin{equation}
H|g\rangle=0. \label{13}%
\end{equation}
This is because the eigenvalue is equal to $\langle g|H|g\rangle$ and the
following relation holds:
\[
\langle g|H|g\rangle=\sum_{n}\langle g|T_{n}|g\rangle=0.
\]
From Eq. (\ref{13}), it is ensured that the Hamiltonian $H$ is a nonnegative operator:%

\begin{equation}
H\geq0. \label{e14}%
\end{equation}

Next, let us define a separable ground state. The separable ground state is
the ground state that takes the form of a product of each-site pure states
such that%

\begin{equation}
|g\rangle=%
%TCIMACRO{\dprod \limits_{n}}%
%BeginExpansion
{\displaystyle\prod\limits_{n}}
%EndExpansion
|g_{n}\rangle. \label{e16}%
\end{equation}
Here $|g_{n}\rangle$ is a pure state for the subsystem at site $n$. This
separable ground state satisfies the factorization property. For instance, a
two-point function of $T_{n}$ and a local operator $O_{m}$ at site $m$ with
$\left\vert n-m\right\vert >1$ is equal to the product of the one-point
functions for the ground state $|g\rangle$:
\begin{equation}
\langle g|T_{n}O_{m}|g\rangle=\langle g|T_{n}|g\rangle\langle g|O_{m}%
|g\rangle. \label{e15}%
\end{equation}
It is well known that the relation in Eq. (\ref{e15}) is often broken for
ordinary quantum systems. This implies that the ground states of standard
many-body systems are usually nonseparable, and do not satisfy Eq.
(\ref{e16}). Such a nonseparable ground state with
\begin{equation}
|g\rangle\neq%
%TCIMACRO{\dprod \limits_{n}}%
%BeginExpansion
{\displaystyle\prod\limits_{n}}
%EndExpansion
|g_{n}\rangle. \label{81}%
\end{equation}
is called an entangled ground state. In the entangled ground state, quantum
fluctuations of subsystems share a quantum correlation, that is, entanglement.
Though entanglement is an informational concept, it is simultaneously a
\textit{physical} resource of quantum communication. For a detailed
explanation, see the text book by \cite{nc}.

If a ground state satisfies the relation of broken factorization,
\begin{equation}
\langle g|T_{n}O_{m}|g\rangle\neq\langle g|T_{n}|g\rangle\langle
g|O_{m}|g\rangle\label{e17}%
\end{equation}
for the energy density at site $n$ and a local operator $O_{m}$ with
$\left\vert n-m\right\vert >1$, then the ground state is entangled. This
ground-state entanglement leads to an interesting result. It can be proven by
use of entanglement that the energy density $T_{n}$ takes a negative value
even though the total Hamiltonian is nonnegative. In order to verify this, let
us first prove a useful lemma: The lemma states that if the entangled ground
state $|g\rangle$ satisfies the relation in Eq. (\ref{e17}), then $|g\rangle$
is not an eigenstate of $T_{n}$. This is because assuming $T_{n}%
|g\rangle=\varepsilon_{n}|g\rangle$ with an eigenvalue $\varepsilon_{n}$ leads
to the factorization in Eq. (\ref{e15}) and contradicts Eq. (\ref{e17}). In
fact, using $\langle g|T_{n}=\varepsilon_{n}\langle g|$ and $\varepsilon
_{n}=\langle g|T_{n}|g\rangle$, we can directly derive Eq. (\ref{e15}) as
follows.
\[
\langle g|T_{n}O_{m}|g\rangle=\varepsilon_{n}\langle g|T_{n}O_{m}%
|g\rangle=\langle g|T_{n}|g\rangle\langle g|O_{m}|g\rangle.
\]
By use of this lemma, we next show that the lowest eigenvalue $\epsilon
_{-}(n)$ of $T_{n}$ is negative. The operator $T_{n}$ can be spectrally
decomposed into%

\[
T_{n}=\sum_{\nu,k_{\nu}}\epsilon_{\nu}(n)|\epsilon_{\nu}(n),k_{\nu}%
,n\rangle\langle\epsilon_{\nu}(n),k_{\nu},n|,
\]
where $\epsilon_{\nu}(n)$ are eigenvalues of $T_{n}$; $|\epsilon_{\nu
}(n),k_{\nu},n\rangle$ are corresponding eigenstates in the total Hilbert
space of the chain system; and the index $k_{\nu}$ denotes the degeneracy
freedom of the eigenvalue $\epsilon_{\nu}(n)$. The ground state can be
expanded as%
\[
|g\rangle=\sum_{\nu,k_{\nu}}g_{\nu,k_{\nu}}(n)|\epsilon_{\nu}(n),k_{\nu
},n\rangle,
\]
where $g_{\nu,k_{\nu}}(n)$ are complex coefficients. Using this expansion, Eq.
(\ref{12}) is rewritten as
\begin{equation}
\sum_{\nu,k_{\nu}}\epsilon_{\nu}(n)\left\vert g_{\nu,k_{\nu}}(n)\right\vert
^{2}=0. \label{e18}%
\end{equation}
If $\epsilon_{-}(n)$ is positive, Eq. (\ref{e18}) clearly has no solution for
$g_{\nu,k_{\nu}}(n)$; thus, it is impossible. If $\epsilon_{-}(n)$ is zero,
then Eq. (\ref{e18}) has a solution with nonvanishing $g_{-,k_{-}}(n)$.
Because all the other coefficients $g_{\nu,k_{\nu}}(n)$ must vanish, this
means that $|g\rangle$ is an eigenstate of $T_{n}$ with $\epsilon_{-}(n)=0$.
Therefore, this contradicts Eq. (\ref{e17}) via the above lemma. Therefore,
$\epsilon_{-}(n)$ must be negative:%
\[
\epsilon_{-}(n)=-\left\vert \epsilon_{-}(n)\right\vert <0.
\]
The average energy density for $|\epsilon_{-}(n),n\rangle$ also becomes
negative. It is thereby verified that there exist quantum states with negative
energy density. It should be stressed that even if a state has negative energy
density over a certain region, there exists compensating positive energy
density in other regions such that the total energy is not negative, because
of the nonnegativity of $H$.

In the later discussion, we adopt general measurements beyond ideal
(projective) measurements. Here, let us give a brief overview of the general
measurements, which are usually called positive operator valued measure (POVM)
measurements. Let us first consider a quantum system $S$ in a state
$|\psi\rangle_{S}$ about which we wish to obtain information. That is, $S$ is
the target system of the measurement. In order to execute quantum
measurements, we need another quantum system $P$ as a probe. Initially, $P$ is
in a state $|0\rangle_{P}$. In general, the dimensionality of the Hilbert
space of $S$ is not equal to that of $P$. We bring $P$ into contact with $S$
via measurement interactions between the two. In this process, information
about $|\psi\rangle_{S}$ is imprinted into $P$. After switch-off of the
measurement interactions and subsequent signal amplification of the probe
system, the total system is in an entangled state that takes a form
\[
|\Psi\rangle_{SP}=\sum_{n,\mu}c_{n\mu}|n\rangle_{S}|\mu\rangle_{P}.
\]
Here, $\left\{  |n\rangle_{S}\right\}  $ is a complete set of orthonormal
basis state vectors of $S$, and $\left\{  |\mu\rangle_{P}\right\}  $ is a set
of orthonormal state of $P$. The coefficient $c_{n\mu}$ depends on the initial
state $|\psi\rangle_{S}$ of $S$. For the state $|\Psi\rangle_{SP}$, a
projective measurement detecting which $|\mu\rangle_{P}$ is realized for $P$
is performed in order to obtain imprinted information about $|\psi\rangle_{S}%
$. This completes a general measurement. The emergence probability of $\mu$ is
given by
\[
p_{\mu}=\sum_{n}|c_{n\mu}|^{2},
\]
which is dependent on $|\psi\rangle_{S}$. Such a general measurement can be
always described using measurement operators $M_{S}(\mu)$ \cite{nc}, which act
on the Hilbert space of $S$ and satisfy%

\[
\sum_{\mu}M_{S}(\mu)^{\dag}M_{S}(\mu)=I_{S},
\]
where the number of $M_{S}(\mu)$ is not generally equal to the number of
dimensions of the Hilbert space of $S$. It should be stressed that in general,
$M_{S}(\mu)$ is not a projective Hermitian operator. It can be shown that for
an arbitrary quantum state $\rho_{S}$ of $S$, the emergence probability
$p(\mu)$ of $\mu$ can be calculated as%

\[
p(\mu)=\operatorname*{Tr}\left[  \rho_{S}M_{S}(\mu)^{\dag}M_{S}(\mu)\right]
.
\]
The post-measurement state of $S$ can be computed as%

\[
\rho(\mu)=\frac{M_{S}(\mu)\rho_{S}M_{A}(\mu)^{\dag}}{\operatorname*{Tr}\left[
\rho_{S}M_{S}(\mu)^{\dag}M_{S}(\mu)\right]  }.
\]
In mathematics, the set of Hermitian positive semidefinite operators
$M_{S}(\mu)^{\dag}M_{S}(\mu)$ is called positive operator valued measure (POVM
for short). This is because the general measurement is often called POVM measurement.

Next, let us construct a QET protocol with a discrete chain system. Let us
assume that Alice stays in front of one subsystem $A$ at $n=n_{A}$, and Bob
stays in front of another subsystem $B$ at $n=n_{B}$ with $\left\vert
n_{A}-n_{B}\right\vert \geq5$. Because they are sufficiently separated from
each other, it is not only the local operators of $A$ but also the energy
density operator $T_{n_{A}}$, which is semi-local, that commute with local
operators of $B$ and $T_{n_{B}}$. At the first step of the QET protocol, Alice
performs a POVM measurement on $A$, which is described by measurement
operators $M_{A}\left(  \alpha\right)  $ with output $\alpha$ satisfying the
sum rules given by
\begin{equation}
\sum_{\alpha}M_{A}\left(  \alpha\right)  ^{\dag}M_{A}\left(  \alpha\right)
=I_{A}. \label{e21}%
\end{equation}
The POVM of this measurement is defined by%
\begin{equation}
\Pi_{A}\left(  \alpha\right)  =M_{A}\left(  \alpha\right)  ^{\dag}M_{A}\left(
\alpha\right)  \label{29}%
\end{equation}
Then emergence probability of $\alpha$ is computed for the ground state as%
\[
p_{A}(\alpha)=\langle g|\Pi_{A}\left(  \alpha\right)  |g\rangle.
\]
The post-measurement state corresponding to $\alpha$ is given by%

\[
|A(\alpha)\rangle=\frac{1}{\sqrt{p_{A}(\alpha)}}M_{A}\left(  \alpha\right)
|g\rangle.
\]
The average post-measurement state is calculated as%

\[
\rho_{M}=\sum_{\alpha}p_{A}(\alpha)|A(\alpha)\rangle\langle A(\alpha
)|=\sum_{\alpha}M_{A}\left(  \alpha\right)  |g\rangle\langle g|M_{A}\left(
\alpha\right)  ^{\dag}.
\]
Therefore, the expectation value of total energy after the measurement is
evaluated as%

\[
E_{A}=\operatorname*{Tr}\left[  H\rho_{M}\right]  =\sum_{\alpha}\langle
g|M_{A}\left(  \alpha\right)  ^{\dag}HM_{A}\left(  \alpha\right)  |g\rangle.
\]
Due to the passivity of $|g\rangle$, $E_{A}$ is positive. Thus, the
measurement device infuses energy $E_{A}$ into the chain system during the
measurement. $E_{A}$ is the input energy of the QET protocol. Because we
consider a short time scale for the QET protocol, time evolution of the chain
system can be neglected. Hence, the input energy $E_{A}$ is localized around
site $n_{A}$ after the measurement. To see this directly, let us introduce a
local energy operator $H_{A}$ around site $n_{A}$ by the sum of energy density
operators that include contributions from $A$:%

\[
H_{A}=\sum_{n=n_{A}-1}^{n_{A}+1}T_{n}.
\]
Let us also define an energy operator outside of $n_{A}$ as
\[
H_{\bar{A}}=H-H_{A.}%
\]
Then $E_{A}$ can be computed as
\begin{align}
E_{A}  &  =\sum_{\alpha}\langle g|M_{A}\left(  \alpha\right)  ^{\dag}%
H_{A}M_{A}\left(  \alpha\right)  |g\rangle\nonumber\\
&  +\sum_{\alpha}\langle g|M_{A}\left(  \alpha\right)  ^{\dag}H_{\bar{A}}%
M_{A}\left(  \alpha\right)  |g\rangle. \label{e20}%
\end{align}
Because of the commutation relation given by
\[
\left[  H_{\bar{A}},~M_{A}\left(  \alpha\right)  \right]  =0,
\]
the second term on the right-hand side in Eq. (\ref{e20}) vanishes as
follows.
\begin{align*}
&  \sum_{\alpha}\langle g|M_{A}\left(  \alpha\right)  ^{\dag}H_{\bar{A}}%
M_{A}\left(  \alpha\right)  |g\rangle\\
&  =\langle g|\left(  \sum_{\alpha}M_{A}\left(  \alpha\right)  ^{\dag}%
M_{A}\left(  \alpha\right)  \right)  H_{\bar{A}}|g\rangle\\
&  =\langle g|H_{\bar{A}}|g\rangle=\sum_{n\notin\left[  n_{A}-1,n_{A}%
+1\right]  }\langle g|T_{n}|g\rangle\\
&  =0.
\end{align*}
Here, we have used Eqs. (\ref{e21}) and (\ref{12}). Therefore, $E_{A}$ is
equal to the average local energy around site $n_{A}$:%

\begin{equation}
E_{A}=\sum_{\alpha}\langle g|M_{A}\left(  \alpha\right)  ^{\dag}H_{A}%
M_{A}\left(  \alpha\right)  |g\rangle. \label{27}%
\end{equation}
It is also verified that the expectation values of $T_{n}$ with $\left\vert
n-n_{A}\right\vert \geq2$ remain exactly zero after the measurement. This
ensures that the input energy $E_{A}$ is stored locally around site $n_{A}$.
At the second step, Alice announces the measurement result $\alpha$ to Bob via
a classical channel. We can assume that the speed of communication is greater
than that of energy diffusion of the system. Thus, time evolution of the
system is omitted. At the third step, Bob performs a local operation
$U_{B}\left(  \alpha\right)  $ dependent on $\alpha$ on $B$. $U_{B}\left(
\alpha\right)  $ is given by%

\begin{equation}
U_{B}\left(  \alpha\right)  =\exp\left[  -i\alpha\theta G_{B}\right]  ,
\label{31}%
\end{equation}
where $G_{B}$ is a local Hermitian operator on $B$ and $\theta$ is a real
constant set such that a positive amount of energy is teleported via QET.
After the operation, the average state of the chain system becomes%

\begin{equation}
\rho_{QET}=\sum_{\alpha}U_{B}(\alpha)M_{A}(\alpha)|g\rangle\langle
g|M_{A}(\alpha)^{\dag}U_{B}(\alpha)^{\dag}. \label{23}%
\end{equation}
The amount of energy extracted from the chain during the operation is given by%

\[
E_{B}=E_{A}-\operatorname*{Tr}\left[  H\rho_{QET}\right]  .
\]
This is the output energy of the QET protocol. Later, let us evaluate $E_{B}$.
Substituting Eq. (\ref{23}) into the above equation yields%

\begin{equation}
E_{B}=E_{A}-\sum_{\alpha}\langle g|M_{A}\left(  \alpha\right)  ^{\dag}%
U_{B}(\alpha)^{\dag}HU_{B}(\alpha)M_{A}\left(  \alpha\right)  |g\rangle.
\label{25}%
\end{equation}
As in the case of $H_{A}$, let us introduce, for convenience, a local energy
operator around site $n_{B}$ as%

\[
H_{B}=\sum_{n=n_{B}-1}^{n_{B}+1}T_{n}.
\]
In addition, the energy operator $H_{\overline{AB}}$ outside $n_{A}$ and
$n_{B}$ is defined as
\[
H_{\overline{AB}}=H-H_{A}-H_{B}.
\]
By this definition, we can derive the following commutation relations because
of operator locality.%
\begin{align}
\left[  H_{\overline{AB}},~M_{A}\left(  \alpha\right)  \right]   &
=0,\label{e24}\\
\left[  H_{\overline{AB}},~U_{B}(\alpha)\right]   &  =0. \label{e25}%
\end{align}
Because the total Hamiltonian $H$ in Eq. (\ref{25}) is given by a sum of
$H_{A},H_{B}$ and $H_{\overline{AB}}$, we obtain the following relation.%

\begin{align}
E_{B}  &  =E_{A}-\sum_{\alpha}\langle g|M_{A}\left(  \alpha\right)  ^{\dag
}U_{B}(\alpha)^{\dag}H_{A}U_{B}(\alpha)M_{A}\left(  \alpha\right)
|g\rangle\nonumber\\
&  -\sum_{\alpha}\langle g|M_{A}\left(  \alpha\right)  ^{\dag}U_{B}%
(\alpha)^{\dag}H_{B}U_{B}(\alpha)M_{A}\left(  \alpha\right)  |g\rangle
\nonumber\\
&  -\sum_{\alpha}\langle g|M_{A}\left(  \alpha\right)  ^{\dag}U_{B}%
(\alpha)^{\dag}H_{\overline{AB}}U_{B}(\alpha)M_{A}\left(  \alpha\right)
|g\rangle. \label{28}%
\end{align}
The second term on the right-hand side of Eq. (\ref{28}) can be computed as
\begin{align*}
-  &  \sum_{\alpha}\langle g|M_{A}\left(  \alpha\right)  ^{\dag}U_{B}%
(\alpha)^{\dag}H_{A}U_{B}(\alpha)M_{A}\left(  \alpha\right)  |g\rangle\\
&  =-\sum_{\alpha}\langle g|M_{A}\left(  \alpha\right)  ^{\dag}U_{B}%
(\alpha)^{\dag}U_{B}(\alpha)H_{A}M_{A}\left(  \alpha\right)  |g\rangle\\
&  =-\sum_{\alpha}\langle g|M_{A}\left(  \alpha\right)  ^{\dag}H_{A}%
M_{A}\left(  \alpha\right)  |g\rangle=-E_{A},
\end{align*}
where we have used an operator locality relation given by $\left[
H_{A},~U_{B}(\alpha)\right]  =0$, the unitary relation of $U_{B}(\alpha)$, and
Eq. (\ref{27}). This term compensates for the first term ($E_{A}$) and has no
contribution. The fourth term on the right-hand side of Eq. (\ref{28})
vanishes as follows.%

\begin{align*}
&  \sum_{\alpha}\langle g|M_{A}\left(  \alpha\right)  ^{\dag}U_{B}%
(\alpha)^{\dag}H_{\overline{AB}}U_{B}(\alpha)M_{A}\left(  \alpha\right)
|g\rangle\\
&  =\sum_{\alpha}\langle g|M_{A}\left(  \alpha\right)  ^{\dag}U_{B}%
(\alpha)^{\dag}U_{B}(\alpha)H_{\overline{AB}}M_{A}\left(  \alpha\right)
|g\rangle\\
&  =\sum_{\alpha}\langle g|M_{A}\left(  \alpha\right)  ^{\dag}H_{\overline
{AB}}M_{A}\left(  \alpha\right)  |g\rangle\\
&  =\langle g|\left(  \sum_{\alpha}M_{A}\left(  \alpha\right)  ^{\dag}%
M_{A}\left(  \alpha\right)  \right)  H_{\overline{AB}}|g\rangle\\
&  =\langle g|H_{\overline{AB}}|g\rangle=\langle g|H|g\rangle-\sum_{n=n_{A}%
-1}^{n_{A}+1}\langle g|T_{n}|g\rangle-\sum_{n=n_{B}-1}^{n_{B}+1}\langle
g|T_{n}|g\rangle\\
&  =0
\end{align*}
Here, we have used Eq. (\ref{e25}), the unitary relation of $U_{B}(\alpha)$,
Eq. (\ref{e24}), Eq. (\ref{e21}), and Eq. (\ref{12}) in this order. Therefore,
$E_{B}$ is equal to the third term on the right-hand side of Eq. (\ref{28}):%

\[
E_{B}=-\sum_{\alpha}\langle g|M_{A}\left(  \alpha\right)  ^{\dag}U_{B}%
(\alpha)^{\dag}H_{B}U_{B}(\alpha)M_{A}\left(  \alpha\right)  |g\rangle.
\]
Because $U_{B}(\alpha)^{\dag}H_{B}U_{B}(\alpha)$ commutes with $M_{A}\left(
\alpha\right)  $ due to the operator locality, we can rewrite $E_{B}$ as%
\begin{equation}
E_{B}=-\sum_{\alpha}\langle g|\Pi_{A}\left(  \alpha\right)  U_{B}%
(\alpha)^{\dag}H_{B}U_{B}(\alpha)|g\rangle\label{30}%
\end{equation}
by using Eq. (\ref{29}). The output energy $E_{B}$ in Eq. (\ref{30}) can
always take a positive value by properly selecting the value of $\theta$ in
Eq. (\ref{31}). In fact, let us consider a case with a small value of
$\left\vert \theta\right\vert $ in which $U_{B}\left(  \alpha\right)  $ can be
approximated as%

\[
U_{B}\left(  \alpha\right)  =\exp\left[  -i\theta\alpha G_{B}\right]
\sim1-i\theta\alpha G_{B}.
\]
Then $E_{B}$ is evaluated as%

\begin{align*}
E_{B}  &  =-\langle g|\left(  \sum_{\alpha}M_{A}\left(  \alpha\right)  ^{\dag
}M_{A}\left(  \alpha\right)  \right)  H_{B}|g\rangle\\
&  +i\theta\langle g|\left(  \sum_{\alpha}\alpha\Pi_{A}\left(  \alpha\right)
\right)  \left[  H_{B},~G_{B}\right]  |g\rangle+O(\theta^{2}).
\end{align*}
The first term in the above equation vanishes because of Eq. (\ref{e21}) and
Eq. (\ref{12}). Therefore, $E_{B}$ is written as%

\[
E_{B}=\theta\langle g|D_{A}\dot{G}_{B}|g\rangle+O(\theta^{2}),
\]
where $D_{A}$ is a Hermitian operator given by%
\[
D_{A}=\sum_{\alpha}\alpha\Pi_{A}\left(  \alpha\right)
\]
and $\dot{G}_{B}$ is a Hermitian operator given by
\[
\dot{G}_{B}=i\left[  H_{B},~G_{B}\right]  =i\left[  H,~G_{B}\right]  .
\]
By this definition, $\dot{G}_{B}$ denotes the time derivative operator of the
Heisenberg operator $G_{B}(t)=e^{itH}G_{B}e^{-itH}$ at $t=0$. Let us introduce
a significant parameter $\eta$ as a two-point function for the ground state,
given by%

\[
\eta=\langle g|D_{A}\dot{G}_{B}|g\rangle.
\]
The reality of $\eta$ ($\eta^{\ast}=\eta$) is guaranteed by an operator
locality relation given by $\left[  D_{A},~\dot{G}_{B}\right]  =0$. Then
$E_{B}$ is simply written as%

\[
E_{B}=\theta\eta+O(\theta^{2}).
\]
As long as $\eta$ does not vanish, $E_{B}$ can take a positive value;
\[
E_{B}\sim\left\vert \theta\eta\right\vert >0
\]
by taking the same sign of $\theta$ as that of $\eta$:
\[
\theta=\operatorname*{sgn}\left(  \eta\right)  \left\vert \theta\right\vert .
\]
Therefore, it is really possible to teleport a positive amount of energy by
this protocol. It is also verified that local energy around site $n_{B}$ takes
a negative value $-E_{B}$ because of local energy conservation.

Maximization of $E_{B}$ in terms of $\theta$ should be independently performed
for each QET model. For example, in a protocol with a qubit chain, we have the
following result \cite{1}-\cite{2}. The qubit chain is composed of qubits
arrayed in one dimension and coupled by nearest-neighbor interactions. In the
model, Alice measures a local observable $\sigma_{A}$ given by a component of
the Pauli operator in the direction of a three-dimensional unit real vector
$\vec{u}_{A}$ such that
\[
\sigma_{A}=\vec{u}_{A}\cdot\vec{\sigma}_{n_{A}}.
\]
The outputs $\alpha$ are its eigenvalue $\pm1$, and projective operators onto
the eigenstates corresponding to $\alpha$ are denoted by $P_{A}(\alpha)$. Then
the input energy is given by%

\[
E_{A}=\sum_{\alpha}\langle g|P_{A}\left(  \alpha\right)  HP_{A}\left(
\alpha\right)  |g\rangle.
\]
The local generator $G_{B}$ of Bob's operation is given by a component
$\sigma_{B}$ of the Pauli operator in the direction of a three-dimensional
unit real vector $\vec{u}_{B}$ such that
\[
G_{B}=\sigma_{B}=\vec{u}_{B}\cdot\vec{\sigma}_{n_{B}}.
\]
Because the eigenvalues of $G_{B}$ are $\pm1$, the square of $G_{B}$ becomes
the identical operation of $B$:
\[
G_{B}^{2}=I_{B}.
\]
Therefore, we can calculate $U_{B}\left(  \alpha\right)  $ without
approximation as
\begin{align}
U_{B}\left(  \alpha\right)   &  =\exp\left[  -i\theta\alpha G_{B}\right]
\nonumber\\
&  =I_{B}\cos\theta-i\alpha\sigma_{B}\sin\theta. \label{70}%
\end{align}
Using Eq. (\ref{70}), $E_{B}$ in Eq. (\ref{30}) is explicitly computed as
\begin{equation}
E_{B}=\frac{\eta}{2}\sin(2\theta)-\frac{\xi}{2}\left(  1-\cos\left(
2\theta\right)  \right)  , \label{71}%
\end{equation}
where $\eta$ is given by
\[
\eta=\langle g|\sigma_{A}\dot{\sigma}_{B}|g\rangle
\]
with $D_{A}=\sigma_{A}$ and $G_{B}=\sigma_{B}$. The constant $\xi$ is defined
as
\[
\xi=\langle g|\sigma_{B}H\sigma_{B}|g\rangle,
\]
and is positive. Maximization of $E_{B}$ in terms of $\theta$ is achieved by
fixing $\theta$ as
\begin{align}
\cos\left(  2\theta\right)   &  =\frac{\xi}{\sqrt{\xi^{2}+\eta^{2}}}%
,\label{c}\\
\sin(2\theta)  &  =\frac{\eta}{\sqrt{\xi^{2}+\eta^{2}}}. \label{s}%
\end{align}
Substituting Eqs. (\ref{c}) and (\ref{s}) into Eq. (\ref{71}) yields the
maximum value of $E_{B}$ such that%
\[
E_{B}=\frac{1}{2}\left[  \sqrt{\xi^{2}+\eta^{2}}-\xi\right]  .
\]
As long as $\eta$ is nonzero, $E_{B}$ becomes positive.

As a significant qubit chain model, let us consider the critical Ising model
with transverse magnetic field. It has energy density at site $n$
\[
T_{n}=-J\sigma_{n}^{z}-\frac{J}{2}\sigma_{n}^{x}\left(  \sigma_{n+1}%
^{x}+\sigma_{n-1}^{x}\right)  -\epsilon,
\]
where $J$ is a positive constant and $\epsilon$ is a real constant satisfying
$\langle g|T_{n}|g\rangle=0$. The total Hamiltonian reads%

\begin{equation}
H=\sum_{n}T_{n}=-J\left[  \sum_{n=-\infty}^{\infty}\sigma_{n}^{z}%
+\sum_{n=-\infty}^{\infty}\sigma_{n}^{x}\sigma_{n+1}^{x}\right]  -E_{g},
\label{cic}%
\end{equation}
where $E_{g}$ is a constant that shifts the eigenvalue of the ground state
$|g\rangle$ to zero. By the standard treatment of the model, we can
analytically evaluate Alice's input energy \cite{1} as%
\[
E_{A}=\frac{6}{\pi}J.
\]
Meanwhile, Bob's output energy is evaluated \cite{1} as
\[
E_{B}=\frac{2J}{\pi}\left[  \sqrt{1+\left(  \frac{\pi}{2}\Delta(\left\vert
n_{A}-n_{B}\right\vert )\right)  ^{2}}-1\right]  ,
\]
where the function $\Delta(n)$ is defined by%
\[
\Delta(n)=-\left(  \frac{2}{\pi}\right)  ^{n}\frac{2^{2n(n-1)}h(n)^{4}%
}{\left(  4n^{2}-1\right)  h(2n)}%
\]
with%
\[
h(n)=\prod_{k=1}^{n-1}k^{n-k}.
\]
When we take a large separation between Alice and Bob ($\left\vert n_{B}%
-n_{A}\right\vert \gg1$), it is straightforwardly verified that the decay of
$E_{B}$ obeys not an exponential but a power law because of the criticality of
this model. In fact, $E_{B}$ takes an asymptotic form of%
\begin{equation}
E_{B}\sim J\frac{\pi}{64}\sqrt{e}2^{1/6}c^{-6}\left\vert n_{B}-n_{A}%
\right\vert ^{-9/2}, \label{80}%
\end{equation}
where the constant $c$ is evaluated as $c\sim1.28$.

Note that the input energy $E_{A}$ is still stored around Alice even after the
last step of the QET protocol. What happens if Alice attempts to completely
withdraw $E_{A}$ by local operations at site $n_{A}$ after the energy
extraction by Bob? If this was possible, the energy gain $E_{B}$ of Bob might
have no cost. However, if so, the total energy of the chain system becomes
equal to $-E_{B}$ and negative. Meanwhile, we know that the total energy must
be nonnegative. Hence, Alice cannot withdraw energy larger than $E_{A}-E_{B}$
only by her local operations. The main reason for Alice's failure is that the
first local measurement of $A$ breaks the ground-state entanglement between
$A$ and all the other subsystems. In particular, in the case with projective
measurements, the post-measurement state is an exact separable state with no
entanglement between $A$ and the other subsystems. If Alice wants to recover
the original state with zero energy, she must recreate the broken
entanglement. However, entanglement generation, in general, needs nonlocal
operations \cite{nc}. Therefore, Alice cannot recover the state perfectly by
her local operations alone. Thus, a residual energy inevitably remains around
$A$ inside the chain system. This interesting aspect poses a related problem
about residual energy of local cooling. Let us imagine that we stop the QET
protocol soon after Alice's measurement, and attempt to completely withdraw
$E_{A}$ by local operations. By the same argument as above, it can be shown
that this attempt never succeeds because the measurement already breaks the
ground-state entanglement. The subsystem $A$ at site $n_{A}$ is entangled with
other subsystems in the ground state with zero energy. This entanglement is
broken by the projective measurement of $A$, when $A$ jumps into a pure state.
The post-measurement state of the chain system is not the ground state but
instead an excited state carrying $E_{A}$. Even if an arbitrary local
operation $U_{A}$ is performed on $A$, the broken entanglement cannot be
recovered and nonvanishing energy remains inside the chain system. For a long
time interval beyond the short-time scale of this protocol, it is actually
possible to extract $E_{A}$ by local operations with the assistance of
dynamical evolution induced by the nonlocal Hamiltonian $H$. However, in the
short time interval we considered, this dynamical evolution is not available.
Therefore, we conclude that the minimum value $E_{r}$ with respect to
short-time local-cooling operations is always positive. In order to make the
argument more concrete, let us consider a general local-cooling operation on
$A$ after Alice's measurement obtaining the measurement result $\alpha$. It is
known \cite{nc} that the operation is generally expressed by the use of
$\alpha$-dependent local Kraus operators $K_{A}(\alpha,\mu)$ at site $n_{A}$
satisfying%
\begin{equation}
\sum_{\mu}K_{A}^{\dag}(\alpha,\mu)K_{A}(\alpha,\mu)=I_{A}. \label{e30}%
\end{equation}
Then the quantum state after this local cooling of $A$ is given by%
\begin{equation}
\rho_{c}=\sum_{\mu,\alpha}K_{A}(\alpha,\mu)M_{A}\left(  \mu\right)
|g\rangle\langle g|M_{A}^{\dag}\left(  \mu\right)  K_{A}^{\dag}(\alpha,\mu).
\label{e31}%
\end{equation}
The minimum value $E_{r}$ of the residual energy in terms of $K_{A}(\alpha
,\mu)$ satisfying Eq. (\ref{e30}) is defined as%

\begin{equation}
E_{r}=\min_{\left\{  K_{A}(\alpha,\mu)\right\}  }\operatorname*{Tr}\left[
\rho_{c}H\right]  . \label{40}%
\end{equation}
For example, evaluation of $E_{r}$ is performed analytically in the critical
Ising spin model in Eq. (\ref{cic}). The result is obtained in \cite{1} and
given by%

\[
E_{r}=\left(  \frac{6}{\pi}-1\right)  J>0.
\]
Alice cannot extract this energy by any short-time local operation, even
though it really exists in front of her. Because of the nonnegativity of $H$,
it is easily noticed that $E_{r}$ is lower bounded by the teleported energy
$E_{B}$.

It is worth noting that energy can be extracted simultaneously via QET from
not only $B$ but also other subsystems if we know the measurement result of
$A$. Alice stays at $n=0$ and performs the measurement of $A$. She announces
the measurement result $\alpha$ to other sites. Then we can simultaneously
extract energy from many distant sites by local unitary operations
$U_{n}(\alpha)$ dependent on $\alpha$ at site $n$ with $\left\vert
n\right\vert \geq5$. This extended protocol is called quantum energy
distribution (QED for short). It is also proven \cite{1} that the energy input
$E_{A}$ during the measurement of $A$ is lower bounded by the sum of the
teleported energies extracted from other distant sites. Therefore,
analogically speaking from the operational viewpoint, the input energy $E_{A}$
is stored in a form that can be compared to a broad oil field. If we are
authorized users who know the 'password' $\alpha$, we are able to
simultaneously extract energy as \textit{oil} from the quantum system, the
\textit{oil field}, at many sites distant from $A$.

In the conventional forms of energy transportation, impurities in the channel
generate heat when the energy carriers pass through the channel. Thus, time
scale of energy transportation becomes the same order of that of heat
generation. Meanwhile, in QET, because it is not energy but classical
information that is sent, the intermediate subsystems along the channel
between the sender (Alice) and the receiver (Bob) are not excited by the
energy carriers of the system during the short time of a QET process. Much
after the transportation, dynamical evolution of the system begins and then
heat is generated. Thus, the time scale for effective energy transportation by
QET is much shorter than that of heat generation. This property is one of the
remarkable advantages of QET. Due to this property, QET is expected to find
use as an energy distribution scheme inside quantum devices that avoids
thermal decoherence and would thus assist in the development of quantum computers.

Because QET is based on the physics of zero-point fluctuations, which are
usually quite small, the amount of teleported energy is generally small.
However, in a practical application of QET for nanodevices, it would be
possible to consider $N$ discrete chains with one-bit measurements as a single
QET channel with $N$-bit information transfer. Then, the amount of teleported
energy is enhanced by the factor $N$. It is trivial that a large amount of
energy can be transported via QET by the use of parallel arrays of many
quantum chains.

\bigskip

\section{QET with Quantum Field}

\bigskip

~

In section 4, we discussed QET protocols with quantum chains that consist of
subsystems discretely arrayed in one dimension. In this section, we treat a
QET protocol with a massless relativistic field $f$ in one dimension as a
continuum. For a detailed analysis, see \cite{6} and \cite{8}. We adopt the
natural unit $c=\hbar=1$. The equation of motion reads%

\begin{equation}
\left[  \partial_{t}^{2}-\partial_{x}^{2}\right]  f=0. \label{e40}%
\end{equation}
This equation can be exactly solved using the light-cone coordinates such that%

\[
x^{\pm}=t\pm x.
\]
Then Eq. (\ref{e40}) is transformed into
\[
\partial_{+}\partial_{-}f=0.
\]
and general solutions of this equation are given by the sum of the left-mover
component $f_{+}\left(  x^{+}\right)  $ and right-mover component
$f_{-}\left(  x^{-}\right)  $:%
\[
f=f_{+}\left(  x^{+}\right)  +f_{-}\left(  x^{-}\right)  .
\]
The canonical conjugate momentum operator of $f\left(  x\right)  =f|_{t=0}$ is
defined by
\[
\Pi(x)=\partial_{t}f|_{t=0}%
\]
and it satisfies the standard commutation relation,
\[
\left[  f\left(  x\right)  ,~\Pi\left(  x^{\prime}\right)  \right]
=i\delta\left(  x-x^{\prime}\right)  .
\]
The left-moving wave $f_{+}\left(  x^{+}\right)  $ can be expanded in terms of
plane-wave modes as%

\[
f_{+}\left(  x^{+}\right)  =\int_{0}^{\infty}\frac{d\omega}{\sqrt{4\pi\omega}%
}\left[  a_{\omega}^{L}e^{-i\omega x^{+}}+a_{\omega}^{L\dag}e^{i\omega x^{+}%
}\right]  ,
\]
where $a_{\omega}^{L}$ $(a_{\omega}^{L\dag})$ is an annihilation (creation)
operator of a left-moving particle and satisfies%

\begin{equation}
\left[  a_{\omega}^{L},~a_{\omega^{\prime}}^{L\dag}\right]  =\delta\left(
\omega-\omega^{\prime}\right)  . \label{c2}%
\end{equation}
The right-moving wave $f_{-}\left(  x^{-}\right)  $ can also be expanded in
the same way using the plane-wave modes. The energy density operator is given
by
\[
\varepsilon(x)=\frac{1}{2}:\Pi(x)^{2}:+\frac{1}{2}:\left(  \partial
_{x}f(x)\right)  ^{2}:,
\]
where $::$ denotes the normal order of creation--annihilation operators for
the plain-wave modes. The Hamiltonian is given by $H=\int_{-\infty}^{\infty
}\varepsilon(x)dx$. The eigenvalue of the vacuum state has been automatically
tuned to be zero due to the normal ordering in $\varepsilon(x)$:
\[
H|0\rangle=0.
\]
The vacuum state also satisfies
\begin{align*}
a_{\omega}^{L}|0\rangle &  =0,\\
\langle0|\varepsilon\left(  x\right)  |0\rangle &  =0.
\end{align*}
Let us introduce the chiral momentum operators as
\[
\Pi_{\pm}(x)=\Pi\left(  x\right)  \pm\partial_{x}f(x).
\]
Then the energy density can be rewritten as
\begin{equation}
\varepsilon\left(  x\right)  =\frac{1}{4}:\Pi_{+}\left(  x\right)  ^{2}%
:+\frac{1}{4}:\Pi_{-}\left(  x\right)  ^{2}:. \label{e51}%
\end{equation}

We perform a QET protocol for the vacuum state $|0\dot{\rangle}$ as follows:
Let us consider a probe system $P$ of a qubit located in a small compact
region $\left[  x_{A-},x_{A+}\right]  $ satisfying $x_{A_{-}}>0$ in order to
detect zero-point fluctuations of $f$. In a manner similar to that of Unruh
\cite{u}, we introduce a measurement Hamiltonian between $f$ and the qubit
such that
\[
H_{m}(t)=g(t)G_{A}\otimes\sigma_{y},
\]
where $g(t)$ is a time-dependent real coupling constant, $G_{A}$ is given by
\begin{equation}
G_{A}=\frac{\pi}{4}+\int_{-\infty}^{\infty}\lambda_{A}(x)\Pi_{+}\left(
x\right)  dx,\label{600}%
\end{equation}
$\lambda_{A}(x)$ is a real function with support $\left[  x_{A-}%
,x_{A+}\right]  $, and $\sigma_{y}$ is the $y$-component of the Pauli operator
of the qubit. Alice stays in the region $\left[  x_{A-},x_{A+}\right]  $. We
assume that the initial state of the qubit is the up state $|+\rangle$ of the
$z$-component $\sigma_{z}$. In the later analysis, we choose a sudden
switching form such that $g(t)=\delta(t)$. After the interaction is switched
off, we measure the $z$-component $\sigma_{z}$ for the probe spin. If the up
or down state, $\left\vert +\right\rangle $ or $\left\vert -\right\rangle $,
of $\sigma_{z}$ is observed, we assign $\alpha=+$ or $\alpha=-$, respectively,
to the measurement result. The measurement is completed at $t=+0$. The time
evolution of this measurement process with output $\alpha$ can be described by
the measurement operators $M_{A}(\alpha)$, which satisfy%

\[
M_{A}(\alpha)\rho M_{A}(\alpha)^{\dag}=\operatorname*{Tr}_{P}\left[  \left(
I\otimes\left\vert \alpha\right\rangle \left\langle \alpha\right\vert \right)
U(+0)\left(  \rho\otimes\left\vert +\right\rangle \left\langle +\right\vert
\right)  U(+0)^{\dag}\right]  ,
\]
where $\rho$ is an arbitrary density operator of the field, the time evolution
operator $U(+0)=\operatorname*{T}\exp\left[  -i\int_{-0}^{+0}H_{m}(t^{\prime
})dt^{\prime}\right]  $ generated by the instantaneous interaction is computed
as $\exp\left[  -iG_{A}\otimes\sigma_{y}\right]  $, and the trace
$\operatorname*{Tr}_{P}$ is taken to the probe system. The measurement
operators $M_{A}(\alpha)$ are evaluated as%

\[
M_{A}(\alpha)=\left\langle \alpha\right\vert \exp\left[  -iG_{A}\otimes
\sigma_{y}\right]  \left\vert +\right\rangle .
\]
Hence, we obtain the explicit expression of $M_{A}(\alpha)$ such that%

\begin{align}
M_{A}(+)  &  =\cos G_{A},\label{401}\\
M_{A}(-)  &  =\sin G_{A}. \label{402}%
\end{align}
For the vacuum state $|0\rangle$, the emergence probability of $\alpha$ is
independent of $\alpha$ and is given by $1/2$ \cite{8}. The post-measurement
states of $f$ for the result $\alpha$ are calculated as%

\begin{equation}
|\psi(\alpha)\rangle=\sqrt{2}M_{A}(\alpha)|0\rangle=\frac{1}{\sqrt{2}}\left(
e^{-\frac{\pi}{4}i}|\lambda\rangle+\alpha e^{\frac{\pi}{4}i}|-\lambda
\rangle\right)  , \label{ps}%
\end{equation}
where $|\pm\lambda\rangle$ are left-moving coherent states defined by
\begin{equation}
|\pm\lambda\rangle=\exp\left[  \pm i\int_{-\infty}^{\infty}\lambda_{A}%
(x)\Pi_{+}\left(  x\right)  dx\right]  |0\rangle. \label{c1}%
\end{equation}
The two states $|\psi(+)\rangle$ and $|\psi(-)\rangle$ are nonorthogonal to
each other with $\langle\psi(+)|\psi(-)\rangle=\langle\lambda|-\lambda
\rangle\neq0$ because this POVM measurement is not projective \cite{nc}. The
expectation value of the Heisenberg operator of energy density $\varepsilon
\left(  x,t\right)  $ for each post-measurement state is independent of
$\alpha$ and given by
\begin{equation}
\langle\psi(\alpha)|\varepsilon\left(  x,t\right)  |\psi(\alpha)\rangle
=\left(  \partial_{+}\lambda_{A}(x^{+}\right)  )^{2}. \label{ie1}%
\end{equation}
Hence, the amount of total average excitation energy is time-independent and
evaluated as
\begin{equation}
E_{A}=\int_{-\infty}^{\infty}\left(  \partial_{x}\lambda_{A}(x\right)
)^{2}dx. \label{700}%
\end{equation}
The average state at time $T$ is expressed as
\[
\rho_{M}=\sum_{\alpha}e^{-iTH}M_{A}(\alpha)|0\rangle\langle0|M_{A}%
(\alpha)^{\dag}e^{iTH}.
\]
It is worth noting that the state $\rho_{M}$ is a strictly localized state
defined by Knight \cite{knight}, because $\rho_{M}$ is locally the same as
$|0\rangle\langle0|$ at $t=T$ and satisfies $\operatorname*{Tr}\left[
\rho_{M}\varepsilon\left(  x\right)  \right]  =0$ for $x\notin$ $\left[
x_{A-}-T,x_{A+}-T\right]  $.

Bob stays in the region $\left[  x_{B-},x_{B+}\right]  $ with zero energy
density and is on Alice's right-hand side:
\[
x_{A+}<x_{B-}.
\]
Alice sends information about the result $\alpha$, to Bob at $t=+0$ at the
speed of light. Bob receives it at $t=T$. It should be stressed that the
positive-energy wave packet generated by the measurement propagates to the
left from Alice and the information about $\alpha$ propagates to the right
from Alice. Therefore, only classical information is sent from Alice to Bob.
The average energy density of quantum fluctuation around Bob remains zero at
$t=T$. Then Bob performs a unitary operation on the quantum field $f$; the
unitary operation is dependent on $\alpha$ and is given by%

\begin{equation}
U_{B}(\alpha)=\exp\left[  i\alpha\theta\int_{-\infty}^{\infty}p_{B}(x)\Pi
_{+}\left(  x\right)  dx\right]  , \label{509}%
\end{equation}
where $\theta$ is a real parameter fixed below and $p_{B}(x)$ is a real
function of $x$ with its support $\left[  x_{B-},x_{B+}\right]  $. After the
operation, the average state of the field $f$ is given by%

\[
\rho_{F}=\sum_{\alpha}U_{B}(\alpha)e^{-iTH}M_{A}(\alpha)|0\rangle
\langle0|M_{A}(\alpha)^{\dag}e^{iTH}U_{B}(\alpha)^{\dag}.
\]
Let us introduce an energy operator localized around the region $\left[
x_{B-},x_{B+}\right]  $ such that $H_{B}=\int_{-\infty}^{\infty}w_{B}\left(
x\right)  \varepsilon\left(  x\right)  dx$. Here, $w_{B}\left(  x\right)  $ is
a real window function with $w_{B}(x)=1$ for $x\in\left[  x_{B-}%
,x_{B+}\right]  $ and it rapidly decreases outside the region. The average
amount of energy around the region is evaluated \cite{8} as%

\[
\operatorname*{Tr}\left[  \rho_{F}H_{B}\right]  =-\theta\eta+\theta^{2}\xi
\]
where $\xi=\int_{-\infty}^{\infty}\left(  \partial_{x}p_{B}(x)\right)  ^{2}dx$ and%

\begin{equation}
\eta=-\frac{4}{\pi}\left\vert \langle0|2\lambda\rangle\right\vert
\int_{-\infty}^{\infty}\int_{-\infty}^{\infty}p_{B}(x)\frac{1}{\left(
x-y+T\right)  ^{3}}\lambda_{A}(y)dxdy.
\end{equation}
By fixing the parameter $\theta$ such that
\[
\theta=\frac{\eta}{2\xi}%
\]
so as to minimize $\operatorname*{Tr}\left[  \rho_{F}H_{B}\right]  $, it is
proven that the average energy around Bob takes a negative value, that is,%

\begin{equation}
\operatorname*{Tr}\left[  \rho_{F}H_{B}\right]  =-\frac{\eta^{2}}{4\xi}<0.
\label{701}%
\end{equation}
During the operation by Bob, the total average energy decreases by
\begin{equation}
E_{B}=\operatorname*{Tr}\left[  \rho_{M}H\right]  -\operatorname*{Tr}\left[
\rho_{F}H\right]  =E_{A}-\operatorname*{Tr}\left[  \rho_{F}H\right]  .
\label{e60}%
\end{equation}
Because average energy density at $t=T$ vanishes except in the region of the
wave packet excited by Alice's measurement and the region of Bob, the
following relation is proven straightforwardly.%
\begin{equation}
\operatorname*{Tr}\left[  \rho_{F}H\right]  =\operatorname*{Tr}\left[
\rho_{F}H_{A}(T)\right]  +\operatorname*{Tr}\left[  \rho_{F}H_{B}\right]  ,
\label{e61}%
\end{equation}
where $H_{A}(T)=\int_{-\infty}^{\infty}w_{A}\left(  x+T\right)  \varepsilon
\left(  x\right)  dx$ and $w_{A}\left(  x\right)  $ is a real window function
for Alice with $w_{A}(x)=1$ for $x\in\left[  x_{A-},x_{A+}\right]  $ and it
rapidly decreases outside the region. The term $\operatorname*{Tr}\left[
\rho_{F}H_{A}(T)\right]  $ in Eq. (\ref{e61}) is the contribution of the
left-moving positive-energy wave packet generated by Alice's measurement. By
virtue of operation locality, it can be proven that the average energy of the
wave packet remains unchanged after Bob's operation:%
\begin{equation}
\operatorname*{Tr}\left[  \rho_{F}H_{A}(T)\right]  =E_{A}. \label{e62}%
\end{equation}
Substituting Eq. (\ref{e61}) with Eq. (\ref{e62}) into Eq. (\ref{e60}) yields%

\[
E_{B}=-\operatorname*{Tr}\left[  \rho_{F}H_{B}\right]  .
\]
According to local energy conservation, the same amount of energy is moved
from the field fluctuation to external systems, including the device executing
$U_{B}(\alpha)$. Therefore, $E_{B}$ is the output energy of this QET protocol.
By using the result in Eq. (\ref{701}), $E_{B}$ can be evaluated as
\[
\bigskip E_{B}=\frac{4\left\vert \langle0|2\lambda\rangle\right\vert ^{2}}%
{\pi^{2}}\frac{\left[  \int_{-\infty}^{\infty}\int_{-\infty}^{\infty}%
p_{B}(x)\frac{1}{\left(  x-y+T\right)  ^{3}}\lambda_{A}(y)dxdy\right]  ^{2}%
}{\int_{-\infty}^{\infty}\left(  \partial p_{B}(x^{\prime})\right)
^{2}dx^{\prime}}.
\]
The operation by Bob simultaneously generates a wave packet with negative
energy $-E_{B}$ that propagates toward the left-side spatial infinity. The
protocol is summarized in a spacetime diagram in figure 15.%
%TCIMACRO{\FRAME{dtbpFU}{3.0649in}{2.3073in}{0pt}{\Qcb{\QTR{tiny}{Figure 15:
%Spacetime diagram of a QET protocol with a quantum field. The measurement  by
%Alice infuses energy and generates a right-moving wavepacket with positive
%energy. The measurement result is transferred to Bob. By use of the result,
%Bob extracts positive energy from the zero-point fluctuation of the field.
%This generates a right-moving wavepacket with negative energy. The law of
%local energy conservation is retained.}}}{}{fig15.eps}%
%width 3.0649in;  height 2.3073in;  depth 0pt;  original-width 7.2627in;
%original-height 5.4518in;  cropleft "0";  croptop "1";  cropright "1";
%filename 'FIG-arXiv-QET-REVIEW/EPS/fig15.eps';file-properties "XNPEU";}}}%
%BeginExpansion
\begin{center}
\includegraphics[
height=2.3073in,
width=3.0649in
]%
{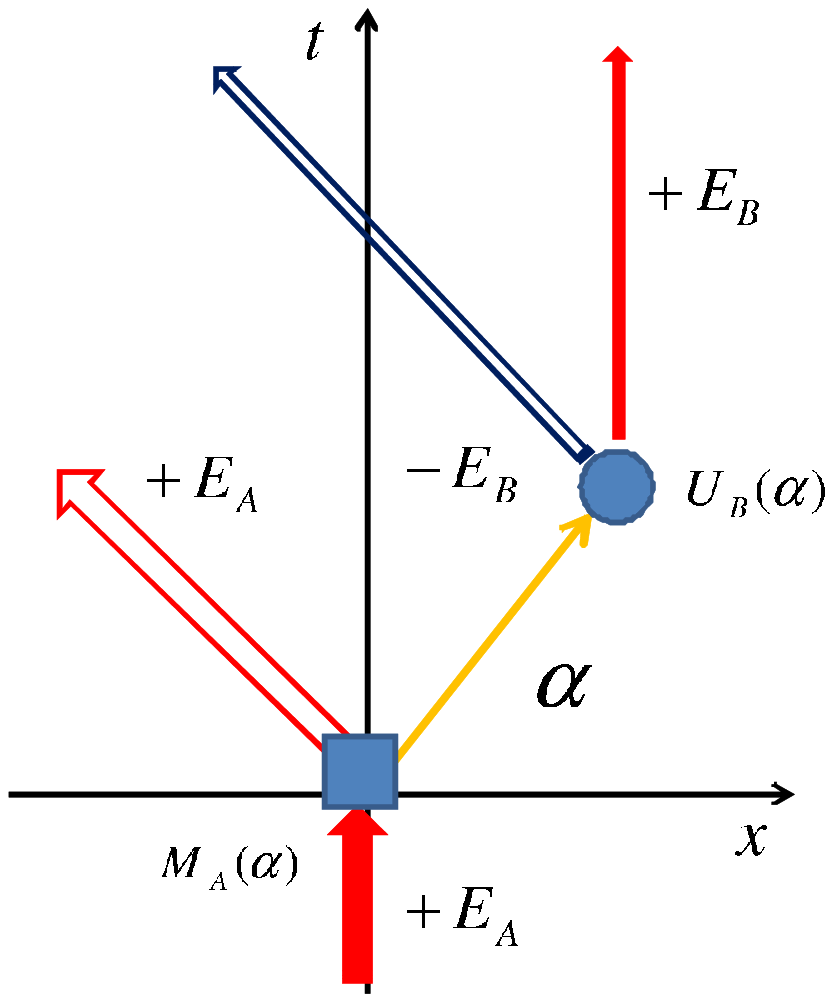}%
\\
{\protect\tiny Figure 15: Spacetime diagram of a QET protocol with a quantum
field. The measurement  by Alice infuses energy and generates a right-moving
wavepacket with positive energy. The measurement result is transferred to Bob.
By use of the result, Bob extracts positive energy from the zero-point
fluctuation of the field. This generates a right-moving wavepacket with
negative energy. The law of local energy conservation is retained.}%
\end{center}
%EndExpansion
Using both this protocol and a previous quantum teleportation protocol in
\cite{QT}, it is possible, in principle, to teleport an object with energy to
a zero-energy local-vacuum region like the above region $\left[  x_{B-}%
,x_{B+}\right]  $.

It is worthwhile to comment that an analogous QET protocol to this protocol
would be experimentally implimented by using quantum Hall edge currents
\cite{YIH}. The edge current can be described by a one-dimensional quantum
scalar field like the system discussed in this section. The most striking
feature of this experimental proposal is that the output energy of the QET
protocol may be of order of $100\mu eV$, which can be observed using current technology.

\bigskip

\section{Summary and Comment}

~

New protocols for quantum energy teleportation (QET) are reviewed that can
accomplish energy transportation by local operations and classical
communication. The protocols do not violate any physical laws, including
causality and local energy conservation. The salient features of QET are
ground-state entanglement of many-body systems and the emergence of negative
energy density due to this entanglement. Research on QET is expected to assist
in the development of quantum nanodevices, including quantum computers. In
addition, QET may shed light on fundamental physics, including quantum
Maxwell's demons, phase transition at zero temperature, and the origin of
black hole entropy.

Finally, a comment is added on energy--entanglement relations in QET. As a
quantitative entanglement measure, negativity is computed between separated
blocks of qubit chains \cite{bose} (the logarithmic negativity for harmonic
oscillator chains \cite{reznik} \cite{5}) showing that at criticality, this
negativity is a function of the ratio of the separation to the length of the
blocks and can be written as a product of a power law and an exponential
decay. This suggests, for the arguments in section 4, that change in the
entanglement between $A$ and $B$ after a local measurement of $A$ has a
similar rapid-decay dependence on the spatial separation. Thus, it may be
concluded that bipartite entanglement between $A$ and $B$ itself is not
essential for QET. Though the bipartite entanglement between the two may be
rapidly damped, $E_{B}$ shows a power law decay ($\propto n^{-9/2}$) for large
spatial separation $n$ at criticality, as seen in Eq. (\ref{80}). In a sense,
this implies that an almost classical correlation between $A$ and $B$ is
sufficient to execute QET for large separation, and is expected to be robust
against environmental disturbances in contrast to the entanglement fragility
in the previous quantum teleportation scheme. It should be emphasized,
however, that this classical correlation is \textit{originally induced by the
ground-state multipartite entanglement} generated by nearest-neighbor
interactions. If the ground state is separable, we have no correlation between
$A$ and $B$. This suggests that teleporting a positive amount of energy
requires some amount of ground-state entanglement. In fact, for the minimal
model of QET discussed in section 3, we have nontrivial energy--entanglement
relations. Let us consider a set of POVM measurements for A which measurement
operators $M_{A}(\mu)$ with measurement output $\mu$ commute with the
interaction Hamiltonian $V~$in the minimal model. These measurements do not
disturb the energy density at B in the ground state. Entropy of entanglement
is adopted as a quantitative measure of entanglement. Before the measurement
of A, the total system is prepared to be in the ground state $|g\rangle$.
\ The reduced state of B is given by $\rho_{B}=\operatorname*{Tr}_{A}\left[
|g\rangle\langle g|\right]  $. The emergent probability $p_{A}(\mu)$ of $\mu$
is given by $\langle g|M_{A}(\mu)^{\dag}M_{A}(\mu)|g\rangle$. After the POVM
measurement outputting $\mu$, the reduced post-measurement state of B is
calculated as $\rho_{B}(\mu)=\frac{1}{p_{A}(\mu)}\operatorname*{Tr}_{A}\left[
M_{A}(\mu)|g\rangle\langle g|M_{A}(\mu)^{\dag}\right]  $. The entropy of
entanglement of the ground state is given by $-\operatorname*{Tr}_{B}\left[
\rho_{B}\ln\rho_{B}\right]  $ and that of the post-measurement state with
output $\mu$ is given by $-\operatorname*{Tr}_{B}\left[  \rho_{B}(\mu)\ln
\rho_{B}(\mu)\right]  $. By using these results, we define the consumption of
ground-state entanglement by the measurement as the difference between the
ground-state entanglement and the averaged post-measurement-state entanglement:%

\[
\Delta S_{AB}=-\operatorname*{Tr}_{B}\left[  \rho_{B}\ln\rho_{B}\right]
-\sum_{\mu}p_{A}(\mu)\left(  -\operatorname*{Tr}_{B}\left[  \rho_{B}(\mu
)\ln\rho_{B}(\mu)\right]  \right)  .
\]
For any measurement which satisfies $\left[  M_{A}(\mu),~V\right]  =0$, the
following relation holds \cite{3}:%

\begin{equation}
\Delta S_{AB}\geq\frac{1+\sin^{2}\varsigma}{2\cos^{3}\varsigma}\ln\frac
{1+\cos\varsigma}{1-\cos\varsigma}\frac{\max E_{B}}{\sqrt{h^{2}+k^{2}}},
\label{ieq}%
\end{equation}
where $\varsigma$ is a real constant fixed by the coupling constants of the
minimal model such that \
\[
\cos\varsigma=\frac{h}{\sqrt{h^{2}+k^{2}}},~\sin\varsigma=\frac{k}{\sqrt
{h^{2}+k^{2}}}.
\]
$\max E_{B}$ is the maximum output energy of QET in terms of the local
operation of B dependent on $\mu$. Eq. (\ref{ieq}) implies that a large amount
of teleported energy really requests a large amount of consumption of the
ground-state entanglement between A and B. It is also noted that for a QET
model with a linear harmonic chain, we have a similar relation between
teleported energy and entanglement \cite{5}. Consequently, it can be said that
the ground-state entanglement really gives birth to QET from the point of view
of information theory \cite{photta}. The ground-state entanglement is a
\textit{physical} resource for energy teleportation.

\bigskip

~

\textbf{Acknowledgments}\newline

I would like to thank Yasusada Nambu for fruitful comments about the
preliminary manuscript. This research has been partially supported by the
Global COE Program of MEXT, Japan, and the Ministry of Education, Science,
Sports and Culture, Japan, under grant no. 21244007.

\end{document}